\documentclass[letter]{article}
\usepackage[utf8]{inputenc}
\usepackage{mathtools}
\usepackage[square,sort,comma,sort&compress,numbers]{natbib}
\usepackage[margin=25mm]{geometry}
\usepackage{graphics}
\usepackage{color}
\usepackage{mathrsfs}
\usepackage{floatrow}   
\usepackage{graphicx,xcolor} 
\usepackage{subfigure}
\usepackage[framemethod=tikz]{mdframed}
\usepackage{amssymb,amsmath,amsfonts,amstext}
\usepackage{stmaryrd}
\usepackage{verbatim}
\usepackage{amsthm}
\usepackage{esint}
\usepackage{caption}
\usepackage{multirow}
\usepackage{bm}
\usepackage{authblk}
\usepackage{cancel}

\graphicspath{ {./Figures/} }

\usepackage{float}

\makeatletter

\providecommand{\keywords}[1]
{
  \small	
  \textbf{\textit{Keywords---}} #1
}

\makeatother

\title{\textbf{Variational Onsager Neural Networks (VONNs): \\
 A thermodynamics-based variational learning strategy for
 non-equilibrium PDEs}}

\author[1]{Shenglin Huang$^\dagger$}
\author[2]{Zequn He$^\dagger$}
\author[1]{Bryan Chem}
\author[1]{Celia Reina\thanks{creina@seas.upenn.edu\\${\color{white}.}\hspace{0.3cm}^\dagger$ Equal contribution }}
\affil[1]{Department of Mechanical Engineering and Applied Mechanics, University of Pennsylvania, Philadelphia, PA 19104, USA}
\affil[2]{Department of Materials Science and Engineering, University of Pennsylvania, Philadelphia, PA 19104, USA}

\begin{document}

\maketitle

\begin{abstract}
We propose a thermodynamics-based learning strategy for non-equilibrium evolution equations based on Onsager's variational principle, which allows to write such PDEs in terms of two potentials: the free energy and the dissipation potential. Specifically, these two potentials are learned from spatio-temporal measurements of macroscopic observables via proposed neural network architectures that strongly enforce the satisfaction of the second law of thermodynamics. The method is applied to three distinct physical processes aimed at highlighting the robustness and versatility of the proposed approach. These include (i) the phase transformation of a coiled-coil protein, characterized by a non-convex free-energy density; (ii) the one-dimensional dynamic response of a three-dimensional viscoelastic solid, which leverages the variational formulation as a tool for obtaining reduced order models; and (iii)
linear and nonlinear diffusion models, characterized by a lack of uniqueness of the free energy and dissipation potentials. These illustrative examples showcase the possibility of learning partial differential equations through their variational action density (i.e., a function instead), by leveraging the thermodynamic structure intrinsic to mechanical and multiphysics problems.

\end{abstract}



\section{Introduction}

The free energy and dissipation potential fully characterize the reversible and irreversible material response under given excitations, including processes such as elasticity, viscoplasticity or reaction-diffusion. Notably, these provide a variational characterization of the partial differential equations governing the non-equilibrium dynamics, potentially far away from equilibrium, through the so-called Onsager's variational principle \citep{onsager1931reciprocala,onsager1931reciprocalb,doi2011onsager,peletier2014variational,arroyo2018onsager,mielke2016generalization}. 
This variational structure not only provides a sense of optimality, which is of marked intellectual beauty, but it is also profoundly powerful from a qualitative and quantitative perspective. Indeed, such variational structure leads to reciprocity of physical interactions, which has been described as `the most fundamental law of Nature revealed up to now' \citep{berdichevsky1989variational}; it leads to evolution equations that are automatically compliant with the second law of thermodynamics (under mild assumptions on the dissipation potential) \citep{mielke2016generalization,arroyo2018onsager}; the symmetries of the action can reveal conserved quantities through the acclaimed Noether's theorem \citep{arnol2013mathematical}; and the principle can further encode stability information that would otherwise not be easily accessible from the equations themselves. From an applications' standpoint, variational principles have also proven extremely powerful for modeling \citep{peletier2014variational,doi2011onsager,arroyo2018onsager}, as a tool for approximations (e.g., reduced order models) \citep{doi2015onsager}, homogenization \citep{dal2012introduction}, variational bounds \citep{hashin1963variational}, numerical methods \citep{glowinski2015variational}, error estimates \citep{radovitzky1999error}, as well as to formulate variational integrators \citep{betsch2016structure}.

Obtaining an accurate characterization of the free energy and the dissipation potential is thus an important scientific quest, which, as noted above, may be seen as more fundamental and insightful than directly discovering the associated partial differential equations. This goal is however endowed with multiple challenges. Firstly, neither the free energy nor the dissipation potential can be measured directly; only some of the derivatives of the free energy are accessible numerically or experimentally. Secondly, statistical mechanics approaches to learn dissipation potentials are only now emerging \citep{montefusco2021framework}, and are currently limited to purely dissipative phenomena (i.e., no coupling to elasticity for instance). Thirdly, their characterization from non-equilibrium data suffers from issues related to sampling of rare events \citep{huang2021harnessing}, which are notoriously difficult, particularly, in an experimental setting. Fourthly and finally, free energy and dissipation potentials are often only computed at specific points, and thus a fitting process is further required to deliver continuous potentials, which is not always trivial \citep{teichert2019machine}.

In this work, we demonstrate that it is possible to directly learn thermodynamic consistent free energy and dissipation potentials from data, while the form of the PDEs naturally follows from Onsager's variational principle and the variables describing the system. For this, we propose neural network architectures, called Variational Onsager Neural Networks (VONNs), that leverage recent advances in machine learning, notably, Integrable Deep Neural Networks (IDNN) \citep{teichert2019machine,teichert2020scale}, Physics Informed Neural Networks (PINNs) \citep{raissi2019physics}, and Fully/Partially Input Convex Neural Networks (FICNN and PICNN) \citep{amos2017input, bunning2021input}. We remark that this approach is in stark contrast with classical literature on PDE and ODE discovery, where the structure of the equations is known \emph{a priori} based on physical knowledge up to certain parameters \citep{raissi2018hidden, raissi2019physics}, or follow (through potentially non-linear functions) from a library of functions/operators and sparsity-promoting techniques. Examples of the latter include, without being comprehensive, sparse identification of nonlinear dynamics (SINDy) \citep{brunton2016discovering}, PDE functional identification of nonlinear dynamics
(PDE-FIND) \citep{rudy2017data}, PDE-Net \citep{long2018pde}, variational system identification (VSI) \citep{wang2019variational} and related strategies \citep{gonzalez1998identification, raissi2018deep, lee2020coarse}.   
The absence of a library of operators leads to minimal restrictions on the form of the resulting PDEs (these are ultimately constrained by the variables used to described the system as will become clearer in the following sections); and greatly reduces human intervention in the setup of the learning strategy.



The proposed approach is also distinct from emerging works on metriplectic dynamical systems that aim at encoding thermodynamic consistency for irreversible processes \citep{ hernandez2021structure, hernandez2021deep, vsipka2021learning,lee2021machine, zhang2021gfinns}. These cited investigations are
nonvariational, restricted to systems governed by quadratic dissipation potentials, and currently limited to describing the evolution by ODEs. More specifically, they are based on the non-variational formulation of the General Equation for Non-Equilibrium Reversible-Irreversible Coupling (GENERIC) formalism \citep{grmela1997dynamics,ottinger2005beyond}, where the evolution equations are written in terms of an energy, an entropy and two operators, which satisfy certain symmetries (or anti-symmetries) as well as degeneracy conditions for thermodynamic consistency. 

We remark that in this investigation, the form of the free energy and dissipation potential densities are fully unconstrained, up to thermodynamic requirements and frame indifference (invariance under the change of observer). It would of course be possible, if so desired, to restrict the functional form of such functions, and reduce the problem to a parameter identification one. Such an inverse problem of material properties identification has been widely treated in the literature, particularly in the context of elastic properties, with strategies such as the constitutive equation gap method (CEGP) \citep{geymonat2002identification} or the virtual fields method (VFM) \citep{pierron2007identification}. A more flexible approach, denoted as variational system identification (VSI) \citep{wang2021inference} aims at non only finding the suitable elastic parameters, but to identify the terms that govern the physics of the problem from a spectrum of admissible ones.

Finally, we note that although the present work combines Onsager's variational principle and PINNs, it is not to be confused with what is noted in the literature as variational PINNs (VPINNs) \citep{kharazmi2021hp}. While we aim at learning the action density of the governing variational principle, the goal of VPINNs is to leverage the weak form of the evolution equations (i.e., the residual is projected onto test functions), to reduce the number of derivatives of the neural networks and hence make the calculations more efficient.


The paper is structured as follows. Section 2 introduces Onsager’s variational principle and discusses its application to inelastic processes in solid mechanics and to diffusion problems. Next, the proposed thermodynamically-consistent neural networks  to learn the free energy and dissipation potential, and the loss function used for training are introduced in Section 3. The resulting learning strategy is then applied to three different models: a phase transformation model in Section 4, a viscoelastic model in Section 5 and linear and nonlinear diffusion models in Section 6. For each of these examples, the model description, data generation, specific network architectures and training results are discussed in detail. Finally, some conclusions and outlooks are provided in Section 7. 


\section{Onsager's variational principle} \label{Sec:OnsagerPrinciple}

For systems at constant temperature, and for which inertia is negligible, Onsager's variational principle reads \citep{onsager1931reciprocala,onsager1931reciprocalb,doi2011onsager,arroyo2018onsager,mielke2016generalization} 

\begin{equation}
     \min_\mathbf{w} \mathcal{R}[\mathbf{z},\mathbf{w}],
\end{equation}
where $\mathcal{R}$ is the Rayleighian, defined as
\begin{equation}
    \mathcal{R}[\mathbf{z},\mathbf{w}]=\dot{\mathcal{F}}[\mathbf{z},\mathbf{w}]+\mathcal{D}[\mathbf{z},\mathbf{w}]+\mathcal{P}[\mathbf{z},\mathbf{w}],
\end{equation}
and $\mathbf{z}$ and $\mathbf{w}$ are, respectively, the state variables and the process variables (the latter describe how the system dissipates energy, and are related to $\dot{z}$ through the so-called process operator). In addition, $\mathcal{F}[\mathbf{z}]$ is the system's free energy, $\mathcal{D}[\mathbf{z},\mathbf{w}]$ is the dissipation potential and $\mathcal{P}[\mathbf{z},\mathbf{w}]$ is the power supplied by the external forces. This variational principle establishes a competition between energy release and dissipation, and can be used to model a wide range of phenomena, including elastic solids undergoing phase transformations \citep{torres2019combined} and viscoplasticity \citep{radovitzky1999error}, phase field models \citep{liu2020variational}, reaction-diffusion systems \citep{mielke2011gradient}, active soft matter \citep{wang2021onsager}, liquid crystals \citep{doi2011onsager}, as well as multiphysics problems that combine several of the above \citep{arroyo2018onsager}. Systems with constraints, such as incompressibility, also exhibit such a variational structure, where the constraints can be naturally added to the variational principle by means of Lagrange multipliers \citep{arroyo2018onsager}.  We refer the reader to \cite{doi2011onsager,peletier2014variational} and \cite{arroyo2018onsager} for a comprehensive review on the subject, including many working examples.

We remark that the equations resulting from Onsager's variational principle 
\begin{equation} \label{Eq:Thermo}
    \frac{\delta \dot{\mathcal{F}}}{\delta \mathbf{w}} + \frac{\delta \mathcal{D}}{\delta \mathbf{w}}+\frac{\delta \mathcal{P}}{\delta \mathbf{w}}=0
\end{equation}
may be seen as fully analogous to the Euler-Lagrange equations, where $\mathcal{F}$ plays the role of the potential energy $V$. Indeed, neglecting the kinetic energy, the Lagrangian reads $L(q)=-V(q)$, where $q$ are the generalized coordinates, and the Euler-Lagrange equations take the form \citep[Chapter 1]{goldstein2011classical}
\begin{equation}
    \cancelto{0}{\frac{d}{dt}\frac{\partial L}{\partial \dot{q}_j}}-\frac{\partial L}{\partial q_j}+\frac{\partial D}{\partial \dot{q}_j}=Q_j.
\end{equation}
Here, $D$ is the so-called Rayleigh's dissipation function, and $Q_j$ are non-conservative generalized forces, often, external forces applied on the system.

For the specific case where $\mathcal{P}=0$ (i.e., no external fores doing work on the system), and mild conditions on the dissipation potential, the free energy $\mathcal{F}$ becomes a Lyapunov function of the dynamics, i.e., $\dot{\mathcal{F}} \leq 0$, in accordance with the second law of thermodynamics. These conditions are: (i) $\mathcal{D}[\mathbf{z},\mathbf{w}]$ is convex on the second argument $\mathbf{w}$, (ii) $\mathcal{D}[\mathbf{z},0]=0$, and (iii) the minimum of $\mathcal{D}$ with respect to $\mathbf{w}$ is zero \citep{mielke2016generalization,arroyo2018onsager}. 

In the presence of inertia, Eqs.~\eqref{Eq:Thermo} can be generalized to
\begin{equation} \label{Eq:ThermoKinetic}
    \frac{d}{dt}\frac{\delta \mathcal{K}}{\delta \mathbf{w}}+\frac{\delta \dot{\mathcal{F}}}{\delta \mathbf{w}} + \frac{\delta \mathcal{D}}{\delta \mathbf{w}}+\frac{\delta \mathcal{P}}{\delta \mathbf{w}}=0,
\end{equation}
where $\mathcal{K}$ is the kinetic energy of the system, as expected from Lagrangian mechanics. These equations may be cast variationally as well in a time discretized setting, as done by \cite{radovitzky1999error}. See as well \cite{kraaij2020fluctuation} for the variational formulation of the evolution equations for non-isothermal processes in the presence of inertia.

This abstract formalism will be put into practice in this paper in the context of solid mechanics and diffusive phenomena, where the notions of state and process variables will be made precise in the following two subsections. For the purpose of this investigation, we will consider that these  variables are measurable though not necessarily controllable through external parameters (such as external force in solid mechanics). Although this theoretical working assumption is not always satisfied in practice, it is common in thermodynamics with internal variables \citep{maugin1999thermomechanics}.
Furthermore, in all the cases considered, it will be assumed for simplicity that, both, the free energy and dissipation potential have a density associated to them, i.e., $\mathcal{F}[\mathbf{z}]=\int_{\Omega} f(\mathbf{z}) \, dV$ and $\mathcal{D}[\mathbf{z}, \mathbf{w}]=\int_{\Omega} \psi(\mathbf{z}, \mathbf{w}) \, dV$, where $\Omega$ is the reference domain. It is precisely these densities that we will aim at learning from trajectory data, so as to fully discover the partial differential equations characterizing the evolution of the system.

\subsection{Onsager's variational principle in solid mechanics} \label{Sec:OnsagerSolidMechanics}
For the purpose of this study, we limit ourselves to isothermal processes in homogenous solids. Following the classical notation in continuum mechanics \citep{gurtin2010mechanics}, we
denote by $\boldsymbol \varphi (\mathbf{X},t)$ 
the deformation mapping at material point $\mathbf{X}$ 
in the reference domain $\Omega$ and time $t$. The velocity and acceleration fields are denoted as $\mathbf{v}(\mathbf{X},t)=\dot{\boldsymbol \varphi}(\mathbf{X},t)$ and $\mathbf{a}(\mathbf{X},t)=\ddot{\boldsymbol \varphi}(\mathbf{X},t)$, 
respectively, and the deformations of the body are locally characterized by the deformation gradient $\mathbf{F}=\nabla \boldsymbol \varphi$.
Furthermore, we will adopt the formalism of irreversible thermodynamics with internal variables \citep{maugin1999thermomechanics} in order to describe inelastic phenomena. These internal variables, denoted by $\mathbf{q}$, may include, for instance, viscous strains in the context of viscoelasticity, or the plastic deformation tensor and hardening variables in the context of plastic solids. Consistent with the notation of the previous section, we then denote by $\mathbf{z}$ the ensemble of state variables, $\mathbf{z}=\left\{\mathbf{F}, \mathbf{q} \right\}$,
while the process variables will here simply correspond to $\mathbf{w}=\left\{\mathbf{v}, \dot{\mathbf{q}}\right\}$.

Next, we follow the general Eqs.~\eqref{Eq:ThermoKinetic}, to obtain the equations governing the evolution of the state variables. First, we recall the various functionals: $\mathcal{K}=\int_{\Omega} \frac{1}{2} \rho |\mathbf{v}|^2 \, dV$, where $\rho$ is the mass density; $\mathcal{F}=\int_{\Omega} f(\mathbf{C}, \mathbf{q}) \, dV$, where the free energy density depends on $\mathbf{F}$ through the right Cauchy-Green tensor $\mathbf{C}=\mathbf{F}^T\mathbf{F}$ to ensure material frame indifference; $\mathcal{D}=\int_{\Omega}\psi(\mathbf{C}, \mathbf{q}, \mathbf{v}, \dot{\mathbf{q}}) \, dV$; 
and $\mathcal{P}=-\int_{\partial \Omega_2} \bar{\mathbf{t}} \cdot \mathbf{v} \, dS - \int_{\Omega} \rho \mathbf{b} \cdot \mathbf{v}\, dV$, where $\bar{\mathbf{t}}$ are the imposed tractions on $\partial \Omega_2$, and $\mathbf{b}$ are the body forces per unit mass. Then, the equations governing the system's evolution naturally follow as
\begin{align}
\label{Eq:equilibrium_solid}
    & \nabla \cdot \left(\frac{\partial f}{\partial \mathbf{F}} \right) + \rho \mathbf{b} = \rho \mathbf{a} + \frac{\partial \psi}{\partial \mathbf{v}}, \quad \text{in} \quad \Omega \\
\label{Eq:BC_solid}
    &\frac{\partial f}{\partial \mathbf{F}} \cdot \mathbf{N} = \bar{\mathbf{t}}, \quad \text{on} \quad \partial \Omega_2 \\
\label{Eq:internal_constitutive_solid}
    & \frac{\partial f}{\partial \mathbf{q}} +\frac{\partial \psi}{\partial \dot{\mathbf{q}}} =0, \quad \text{in} \quad \Omega,
\end{align}
where $\mathbf{N}$ 
is the outward normal to the reference domain. These equations, are the well-known equilibrium equations (in the interior and boundary of the domain) and evolution equations of the internal variables. Although the term $\partial_{\mathbf{v}} \psi$ in Eq.~\eqref{Eq:equilibrium_solid} is not standard in the theory of elasticity, it is naturally occurring in solids subjected to viscous drag (an example of this will be shown in Section \ref{Sec:PhaseTransformation}). As for Eqs.~\eqref{Eq:internal_constitutive_solid}, these may be equivalently found in the literature as $\mathbf{y} = \partial_{\dot{\mathbf{q}}} \psi$, with $\mathbf{y}=-\partial_{\mathbf{q}} f$ being the thermodynamic forces conjugate to the internal variables. Equations \eqref{Eq:equilibrium_solid}-\eqref{Eq:internal_constitutive_solid} in a time discrete setting, are also often denoted as variational constitutive updates in the computational mechanics community \citep{ortiz1999variational}, where it has proven useful for error estimation and for implementing adaptative meshing strategies \citep{radovitzky1999error}.

Finally, we note that the conditions set on the dissipation potential density for thermodynamic consistency are also necessary to uniquely gather the free energy and dissipation potential, at least for problems with a single scalar internal variable $q$.
Indeed, without the condition $\psi(\mathbf{z},0)=\partial_{\mathbf{w}} \psi(\mathbf{z},0)=0$, it would be possible to add and substract an arbitrary constant $C$ to Eq.~\eqref{Eq:internal_constitutive_solid}, hence arbitrarily modifying the dissipation potential density by $C\dot{q}$ and the free energy density by $-Cq$. 




\subsection{Onsager's variational principle for diffusion processes}
\label{Sec:Onsager_Diffusion}

We now turn our attention to diffusive processes in a given domain $\Omega$, and denote by $c(\mathbf{X}, t)$ the concentration of the diffusive substance of interest at a spacial point $\mathbf{X}$ and time $t$. By local conservation, $\dot{c} \coloneqq \frac{\partial c}{\partial t}$ is related to the flux $\mathbf{j}$ as
\begin{equation}
\label{Eq:MassConservation}
    \dot{c} = - \nabla \cdot \mathbf{j}.
\end{equation}
For simplicity, we will consider that the boundary of the domain is impermeable, i.e., $\mathbf{j}\cdot{\mathbf{N}}=0$, or that the simulation domain is characterized by periodic boundary conditions. 

For this class of problems, the state of the system is fully characterized by the concentration field, i.e., $z=c$, while dissipation is naturally associated to the flux $\mathbf{j}$ or some function of $\mathbf{j}$ (such as the ratio $\mathbf{j}/c$ for the classical diffusion problem \citep{arroyo2018onsager}). Without loss of generality, we choose the flux as the process variables, i.e., $\mathbf{w}=\mathbf{j}$, and hence Eq.~\eqref{Eq:MassConservation} characterizes the process operator introduced earlier.

We may now follow Eq.~\eqref{Eq:Thermo}, with $\mathcal{F}=\int_{\Omega}f(c) \, dV$, $\mathcal{D}=\int_{\Omega} \psi(c,\mathbf{j})\, dV$, and $\mathcal{P}=0$ to deliver
\begin{equation}
\label{Eq:DiffKinetic}
    \nabla f'(c) + \frac{\partial \psi(c, \mathbf{j})}{\partial \mathbf{j}} = \mathbf{0},
\end{equation}
which represents the constitutive relation for the flux $\mathbf{j}$ as a function of $c$ and $\nabla c$. This equation, combined with the mass conservation law Eq.~\eqref{Eq:MassConservation}, fully characterizes the evolution of the concentration.

As discussed in the general setting, thermodynamic consistency requires, assuming sufficient smoothness, that $\psi(c, 0) = 0$ and $\partial_{\mathbf{j}} \psi(c, 0) = 0$. Such requirements are also useful to avoid modifying $f(c)$ and $\psi(c, \mathbf{j})$ by an arbitrary additive function $g(c)$, similarly to what occurred in the equations for solid mechanics. However, it is important to note that even with these thermodynamic constraints and the physically-intuitive condition $f(0)=0$, the uniqueness of $f$ and $\psi$ is still not guaranteed. As a counterexample, we remark the existence of two distinct physical processes, namely a zero-range process (ZRP) and the symmetric simple exclusion process (SSEP), that have distinct free energy and dissipation potential densities, while they are both governed by the same macroscopic diffusion equation, $\dot{c} = \nabla \cdot \left( D \nabla c \right)$, where $D$ is the diffusivity. More specifically, the potential densities for the ZRP considered are given by $f_{ZRP}(c) = \beta^{-1}\left[ c \log c-c \right]$ and $\psi_{ZRP}(c,\mathbf{j}) = \left\| \mathbf{j} \right\|^2 / (2 D \beta c)$, while these two functions for the SSEP are given by $f_{SSEP} (c) = \beta^{-1} \left[ c \log c + (1-c) \log (1-c) \right]$ and $\psi_{SSEP}(c, \mathbf{j}) =  \left\| \mathbf{j} \right\|^2 / [2 D \beta c (1-c)]$. For a review of the ZRP and SSEP, we refer the reader to \cite{embacher2018computing} Section 4(a) with $g(k)=k$ and Section 4(b), respectively.

While the lack of uniqueness just discussed implies that it is impossible to recover the physical functions $f$ and $\psi$ from spatio-temporal data on $c$ and $\mathbf{j}$ alone, one may still uniquely recover the constitutive relation for $\mathbf{j}$ and hence, the macroscopic evolution equation for the concentration. Actually, Eq.~\eqref{Eq:DiffKinetic} may be equivalently rewritten by means of a single function $\hat{\psi}(c,\mathbf{j})\coloneqq\psi(c,\mathbf{j})/f''(c)$ 
as
\begin{equation}
\label{Eq:DiffKinetic_psi_tilde}
    \frac{\partial \hat{\psi}(c, \mathbf{j})}{\partial \mathbf{j}} 
    = - \nabla c. 
\end{equation}
For both the ZRP and SSEP above described, $\hat{\psi} = \|\mathbf{j}\|^2 / (2D)$ and thus, they both follow the same constitutive relation $\mathbf{j} = - D \nabla c$ and evolution equation for $c$.

\section{Variational Onsager Neural Networks (VONNs)}

\begin{figure}[H]
    \centering
    \includegraphics[width=\textwidth]{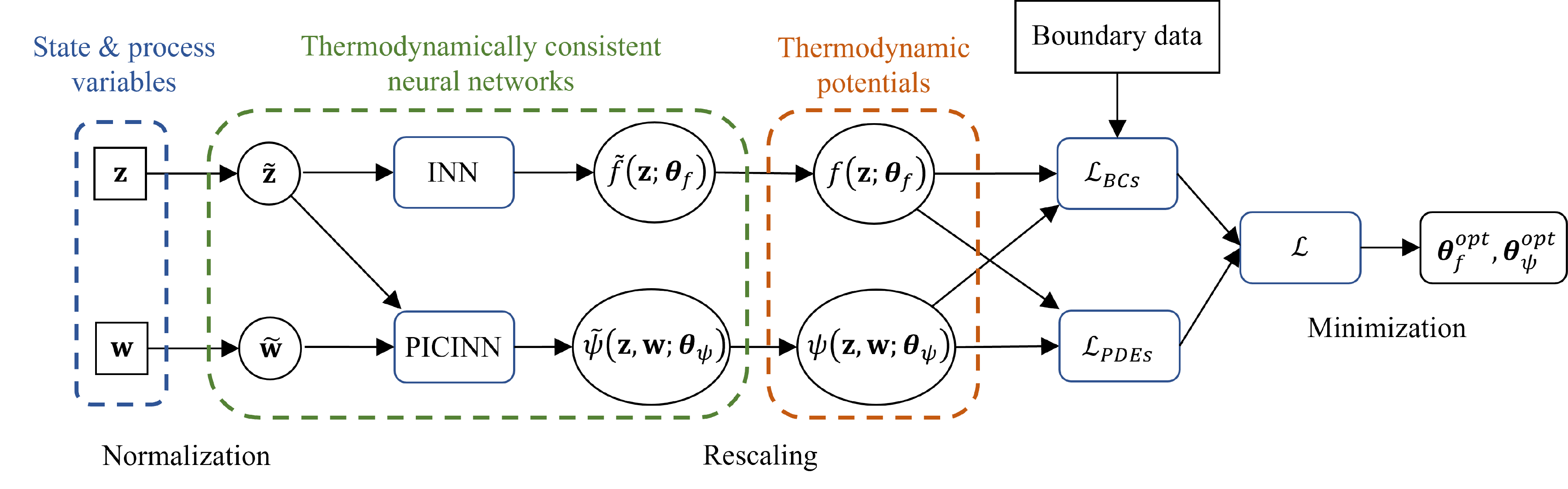}
    \caption{Schematic of the general architecture used to learn the free energy and dissipation potential densities, $f$ and $\psi$, from spatio-temporal data of the state and process variables, $\mathbf{z}$ and $\mathbf{w}$, as well as boundary data, if applicable.}
    \label{Fig:Architecture}
\end{figure}

We here describe the general trainable architecture used to learn the free energy and dissipation potential densities for a given process from spatio-temporal data of the state and process variables, as well as boundary data, if applicable. As schematically shown in Fig.~\ref{Fig:Architecture}, the free energy density and the dissipation potential density are each represented by an independent neural network, and their corresponding training parameters (weights and biases), $\boldsymbol \theta_{f}$ and $\boldsymbol \theta_{\psi}$, are learned by minimizing a loss function in the spirit of Physics-Informed Neural Network (PINNs). Here, this loss corresponds to the $L^2$ norm squared of the residual of the governing equations ($\mathcal{L}_{PDEs}$) and boundary conditions ($\mathcal{L}_{BCs}$), which may be naturally obtained from the free energy, dissipation potential and external power, by means of Onsager's variational principle, as discussed in Section~\ref{Sec:OnsagerPrinciple}. Since only partial derivatives of the $f$ and $\psi$ participate in the training process, their associated neural networks are denoted in the literature as Integrable Neural Networks (INN), or Integrable Deep Neural Networks (IDNN) when the number of hidden layers is larger than one \citep{teichert2019machine}. For the case of the dissipation potential, we will further strongly impose the convexity condition with respect to $\mathbf{w}$ in the neural network by means of Partially Input Convex Integrable Neural Network (PICINN), or a Fully Input Convex Integrable Neural Network (FICINN) when $\psi$ is independent of $\mathbf{z}$ \citep{amos2017input, bunning2021input}. We remark that no convexity restrictions will be imposed on the free energy, as one of our applications of interest will be on a system exhibiting a phase transformation, which is precisely characterized by a non-convex free energy density. The specificities of the individual networks as well as the strategy used to further impose further physical conditions, such as $\psi(\mathbf{z},0)=0$, $\partial_{\mathbf{w}}\psi(\mathbf{z},0)=0$, or $f(0)=0$, are discussed in detail in the following subsections.

Finally, we note that, as is customary in the PINNs literature, the inputs and outputs of the neural networks are normalized and rescaled, respectively, to ensure that both, the input and output, are approximately of scale one -- this is known to facilitate the learning task and improve the robustness of the neural networks \citep{kissas2020machine}.  Specifically, the state and process variables $\mathbf{z}$ and $\mathbf{w}$ are normalized to $\tilde{z}_i = \left( z_i - \mu_{z_i} \right) / \sigma_{z_i} $ and $ \tilde{w}_i = \left( w_i - \mu_{w_i} \right) / \sigma_{w_i} $ for each component $i$, where $\mu_{z_i}$, $\mu_{w_i}$ and $\sigma_{z_i}$, $\sigma_{w_i}$ are the means and standard deviations of the state and process variables, respectively. The rescaling factors are obtained from dimensional analysis, and are hence problem specific. The specific rescaling strategy is detailed for each example analyzed in the corresponding sections.

 \begin{figure}[H]
    \centering
    \includegraphics[width=\textwidth]{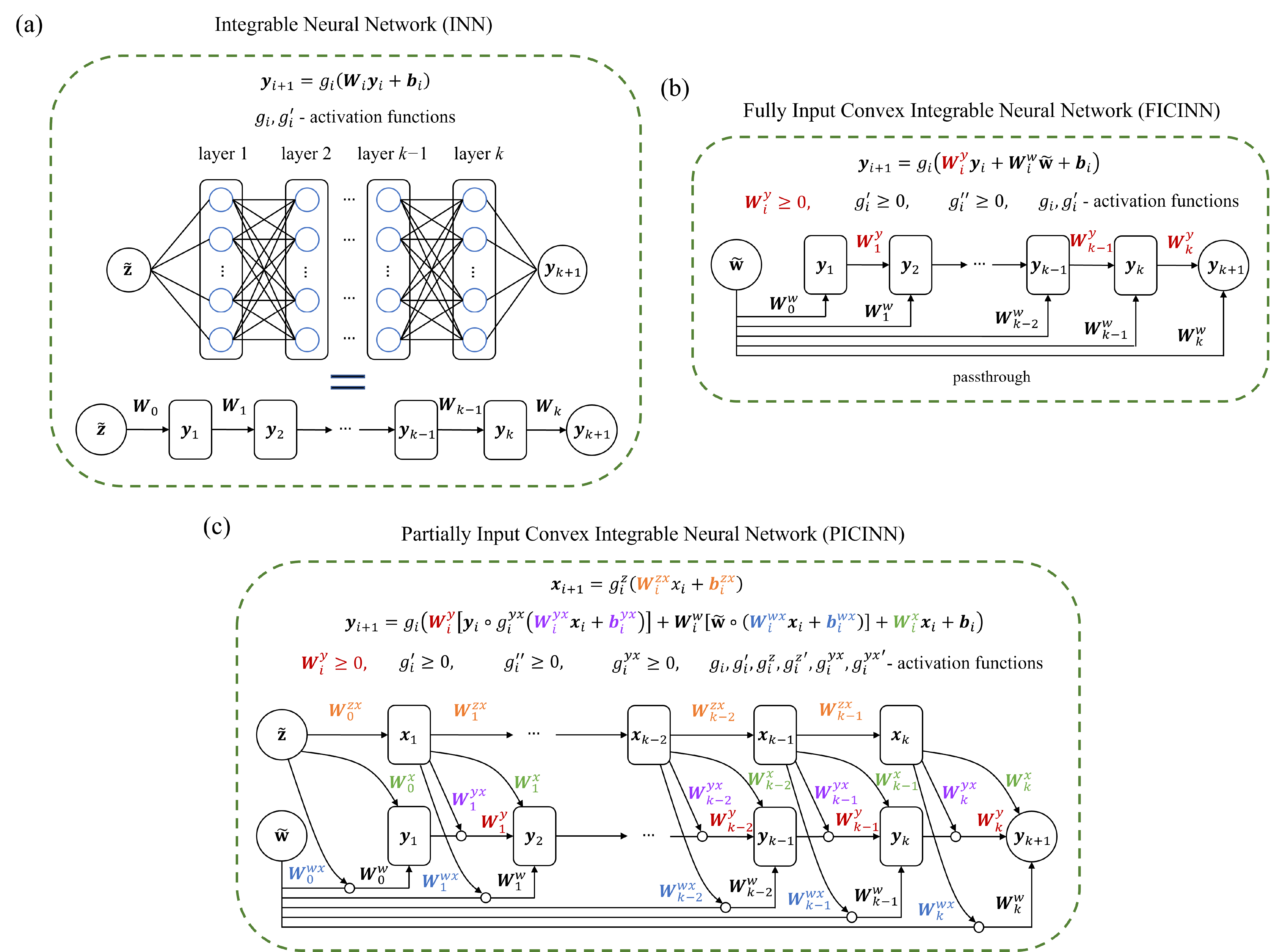}
    \caption{Schematics of the (a) Integrable Neural Network (INN), (b) Fully Input Convex Integrable Neural Network (FICINN) and (c) Partially Input Convex Integrable Neural Network (PICINN).}
    \label{Fig:NN}
\end{figure}

    


\subsection{Free energy density}
\label{Sec:NN_fe}
\noindent

The non-dimensional free energy density $\tilde{f}(\tilde{\mathbf{z}})$ (i.e., where both the input and output are normalized to be of order one) is represented by an Integrable (Deep) Neural Network (INN). The general architecture of the INN with $k$ hidden layers is depicted in Fig.~\ref{Fig:NN}(a). Mathematically, the activation values $\boldsymbol{y}_{i+1}$ of each layer $i+1$ can be expressed as
\begin{equation}
\label{Eq:INN}
    \boldsymbol{y}_{i+1} = g_i \left( \boldsymbol{W}_i \boldsymbol{y}_i + \boldsymbol{b}_i \right), \quad i=0, \dots k,
\end{equation}
where $\boldsymbol{W}_i$, $\boldsymbol{b}_i$ and $g_i(\cdot)$ are the  weights, biases and activation function, and  $\boldsymbol{y}_0 = \tilde{\mathbf{z}}$. We note that $\boldsymbol{y}_i$ and $\boldsymbol{b}_i$ are vectors, while $\boldsymbol{W}_i$ is a matrix, and that $\boldsymbol{W}_i \boldsymbol{y}_i$ denotes a standard matrix vector multiplication. 

A key aspect of INNs, in contrast to regular NNs, is that the activation function $g_i(\cdot)$ should be chosen so that its derivative $g_i'(\cdot)$ is also a common activation function. In this paper, we use the SoftPlus activation function $g_i(x) = \log \left( 1 + e^x \right)$ for all layers in the INN, as in \cite{teichert2019machine}. This activation function is infinitely differentiable, and its derivative is also a common activation function, namely, the logistic function, $g_i'(x) = 1 / \left( 1 + e^{-x} \right)$.

\sloppy

In general, the free energy density $f(\mathbf{z})$ may then be simply recovered from the output of INN $\tilde{f}(\mathbf{z})$\footnote{$\tilde{f}(\mathbf{z})$ is defined, with slight abuse of notation, as the composition of $\tilde{f}(\tilde{\mathbf{z}})$ with $\tilde{\mathbf{z}}(\mathbf{z})$ defined component-wise as $\tilde{z}_i = \left( z_i - \mu_{z_i} \right) / \sigma_{z_i} $.} as ${f(\mathbf{z}) = f^*\tilde{f}(\mathbf{z})} $, where $f^*$ is the characteristic scale of $f(\mathbf{z})$. In many physical problems though, the free energy density is expected to vanish at $\mathbf{z}=0$. In such cases, this condition may be strongly imposed  by defining $f(\mathbf{z})$ as
\begin{equation}
\label{Eq:rescale_fe}
    f(\mathbf{z}) = f^*
    \left[ \tilde{f}(\mathbf{z}) 
    - \tilde{f}(\mathbf{0}) \right].
\end{equation}
Similarly, in other problems, the derivative of free energy with respect to $\mathbf{z}$ or some components of $\mathbf{z}$ is further expected to vanish at the reference state $\mathbf{z}=0$. This is the case, for instance, in the context of viscoelasticity, where the stresses are expected to be zero at zero value of the strain $\boldsymbol \varepsilon$ and viscous strain  $\boldsymbol \varepsilon^v$, i.e., $\partial_{\pmb{\varepsilon}} f(\pmb{\varepsilon} = \mathbf{0}, \pmb{\varepsilon}^v = \mathbf{0}) = \mathbf{0}$. Such condition may as well be strongly enforced by defining the free energy density $f(\boldsymbol \varepsilon,\boldsymbol \varepsilon^v)$ as
\begin{equation}
\label{Eq:rescale_fe_viscoelastic}
    f(\pmb{\varepsilon}, \pmb{\varepsilon}^v) = f^*
    \left[ \tilde{f}(\pmb{\varepsilon}, \pmb{\varepsilon}^v) 
    - \tilde{f}(\mathbf{0}, \mathbf{0})
    - \left. \frac{\partial \tilde{f}}{\partial \boldsymbol \varepsilon} \right|_{\boldsymbol \varepsilon = \mathbf{0}, \pmb{\varepsilon}^v = \mathbf{0}} : \boldsymbol \varepsilon  \right].
\end{equation}
Here, the symbol $:$ denotes the contraction with respect to both indices; that is, for two general second order tensors $\mathbf{A}$ and $\mathbf{B}$, $\mathbf{A}:\mathbf{B}\coloneqq A_{ij}B_{ij}$, where Einstein summation convention is applied.


\subsection{Dissipation potential density}
\label{Sec:NN_psi}
\noindent

Thermodynamic consistency of the resulting continuum equations requires, as discussed in Section \ref{Sec:OnsagerPrinciple}, that (i) $\psi(\mathbf{z}, \mathbf{w})$ is convex with respect to the process variables $\mathbf{w}$, (ii) $\psi(\mathbf{z}, \mathbf{0})=0$, and (iii) $\min_{\mathbf{w}} \psi(\mathbf{z}, \mathbf{w})=0$. This last condition may be equivalently expressed as $\partial_{\mathbf{w}} \psi(\mathbf{z}, \mathbf{w} = \mathbf{0}) = 0$ for smooth functions $\psi$, in view of conditions (i) and (ii). 


To ensure the convexity condition (i) with respect to $\mathbf{w}$, we utilize a Partially Input Convex Neural Network (PICNN), or a Fully Input Convex Neural Network (FICNN) when the disspation potential density is only a function of the process variables $\mathbf{w}$ \citep{amos2017input, bunning2021input}.  We remark that the loss function may still be nonconvex with respect to the parameters of the network $\boldsymbol \theta_{\psi}$, hence making the training problem nonconvex. Moreover, since only the derivatives of the dissipation potential density enter the training process, as occurred with the free energy density, we combine the notions of PICNN/FICNN with that of INN, and hence call these architectures as Partially Input Convex Integrable Neural Networks (PICINNs) and Fully Input Convex Integrable Neural Networks (FICINNs), respectively.


We first describe in detail the simpler case of FICINN to represent $\psi = \psi(\mathbf{w})$. Its architecture with $k$ hidden layers is depicted in Fig.~\ref{Fig:NN}(b) and consists of a traditional feed-forward network (though without the connection between the input layer and the first hidden layer), combined with ``passthrough'' layers that connect the input layer with the units in the following layers. Mathematically, the activation values $\boldsymbol{y}_{i+1}$ (vector) of layer $i+1$ are described by the following relation
\begin{equation}
\label{Eq:FICINN}
    \boldsymbol{y}_{i+1} = g_i \left( \boldsymbol{W}_i^y \boldsymbol{y}_i + \boldsymbol{W}_i^w \tilde{\mathbf{w}} + \boldsymbol{b}_i \right), \quad i=0, \dots k,
\end{equation}
where $\boldsymbol{W}_i^y$ and $\boldsymbol{W}_i^w$ are the weight matrices, $\boldsymbol{b}_i$ are the bias vectors and $g_i(\cdot)$ the activation functions. Furthermore, $\boldsymbol{y}_0 = \tilde{\mathbf{w}}$ is the normalized input, and $\boldsymbol{W}_0^y = \mathbf{0}$. Here, the convexity with respect to $\mathbf{w}$ is guaranteed by using non-negative weights $\boldsymbol{W}_i^{y}$, for $i=1, \dots, k$, and activation functions $g_i(\cdot)$ that are convex and non-decreasing \citep{amos2017input}. These conditions for $g_i(\cdot)$ naturally result from the fact that the sum (with non-negative weights) of convex functions is convex, and that the composition of a convex function and a convex and non-decreasing function is convex \citep{boyd2011distributed}. Furthermore, learning $\psi$ from its derivatives requires that both $g_i(\cdot)$ and $g_i'(\cdot)$ are common activation functions. Here, we utilize the SoftPlus activation function for all layers, which is smooth, convex, non-decreasing and its derivative is also a common activation function. 


For the general case where the dissipation potential density depends on both $\mathbf{z}$ and $\mathbf{w}$, we utilize a PICINN to ensure convexity with respect to $\mathbf{w}$ only. The architecture for the PICINN with $k$ hidden layers is shown in Fig.~\ref{Fig:NN}(c). This may be viewed as a combination of an INN and a FICINN,  with a complex feed-forward structure from each layer of the INN to the following layer in the FICINN. This architecture can be mathematically defined by the following recurrence relation
\begin{equation}
\label{Eq:PICINN}
\begin{split}
    & \boldsymbol{x}_{i+1} = g_i^x \left( \boldsymbol{W}_i^{zx} \boldsymbol{x}_i + \boldsymbol{b}_i^{zx} \right),
    \quad \quad 
    i = 0, 1, \dots, k-1 \\
    & \boldsymbol{y}_{i+1} = g_i \left( 
    \boldsymbol{W}_i^y \left[ \boldsymbol{y}_i \circ 
    g_i^{yx} \left( \boldsymbol{W}_i^{yx} \boldsymbol{x}_i + \boldsymbol{b}_i^{yx} \right) \right]
    + \boldsymbol{W}_i^w \left[ \tilde{\mathbf{w}} \circ 
    \left( \boldsymbol{W}_i^{wx} \boldsymbol{x}_i + \boldsymbol{b}_i^{wx} \right) \right]
    + \boldsymbol{W}_i^x \boldsymbol{x}_i + \boldsymbol{b}_i \right),
    \quad \quad 
    i = 0, 1, \dots, k,
\end{split}
\end{equation}
where $\boldsymbol{x}_{i}$ and $\boldsymbol{y}_{i}$ are the vectors of activation values for layer $i$ in the non-convex portion and convex portion, respectively, $\boldsymbol{W}_i^{zx}$, $\boldsymbol{W}_i^y$, $\boldsymbol{W}_i^{yx}$, $\boldsymbol{W}_i^w$, $\boldsymbol{W}_i^{wx}$ and $\boldsymbol{W}_i^x$ are weight matrices, $\boldsymbol{b}_i^{zx}$, $\boldsymbol{b}_i^{yx}$, $\boldsymbol{b}_i^{wx}$ and $\boldsymbol{b}_i$ are bias vectors, and $g_i^{x}(\cdot)$, $g_i(\cdot)$ and $g_i^{yx}(\cdot)$ are the activation functions for layer $i$. 
Furthermore, $\boldsymbol{x}_0 = \tilde{\mathbf{z}}$ and $\boldsymbol{y}_0 = \tilde{\mathbf{w}}$ are the normalized inputs, $\boldsymbol{W}_0^y=\mathbf{0}$, and the symbol $\circ$ represents the Hadamard product or element-wise product, depicted as well in Fig.~\ref{Fig:NN}(c). Here, the convexity of PICINN with respect to $\mathbf{w}$ is ensured by setting the weights $\boldsymbol{W}_i^y$ to be non-negative, the activation functions $g_i(\cdot)$ to be convex and non-decreasing and the function $g_i^{yx}(\cdot)$ to be non-negative. These conditions have been proposed in literature \citep{amos2017input, bunning2021input}, though the proof is there not provided. For completeness, such a proof is given in Appendix \ref{App:Convex_FICINN}. Similarly to the FICINN, learning $\psi$ from its derivatives requires that $g_i(\cdot)$, $g_i'(\cdot)$, $g_i^{x}(\cdot)$, ${g_i^x}'(\cdot)$, $g_i^{yx}(\cdot)$ and ${g_i^{yx}}'(\cdot)$ are common activation functions. Based on the requirements discussed above, we choose the activation functions
to be the SoftPlus function.

Both, the FICINNs and the PICINNs just introduced require some weights $W$ to be non-negative. From a practical perspective, this is ensured by applying a trick similar to that of \cite{sivaprasad2020curious}, where $W$ is defined as a function of $\tilde{W}$ as
\begin{equation}
    W = 
    \begin{cases}
        \widetilde{W} + \exp(-\epsilon),
        &\widetilde{W} \geq 0\\
        \exp(\widetilde{W}-\epsilon),
        &\widetilde{W} < 0.
    \end{cases}
\end{equation}
Here, $\epsilon$ is a positive constant, which is chosen as $\epsilon=5$ in this work \citep{sivaprasad2020curious}, and $\tilde{W}$ is allowed to take any real value. These auxiliary weights $\tilde{W}$ are then chosen as the trainable parameters for the neural networks instead of $W$. 


Finally, thermodynamic conditions (ii) and (iii), and the rescaling of the FICINN/PICINN output $\tilde{\psi}$, can be jointly considered, by defining the dissipation potential density $\psi$ as
\begin{equation}
\label{Eq:rescale_psi}
    \psi(\mathbf{z}, \mathbf{w}) 
    = \psi^* \left[ 
    \tilde{\psi}(\mathbf{z}, \mathbf{w}) 
    -  \tilde{\psi}(\mathbf{z}, \mathbf{0}) 
    - \left. \frac{\partial\tilde\psi}{\partial \mathbf{w}}\right|_{\mathbf{w}=0} \cdot \mathbf{w} \right],
\end{equation}
where $\psi^*$ is the characteristic scale of $\psi$. It is worth noting that the linear terms in $\mathbf{w}$ added to $\tilde{\psi}$ do not affect its convexity, and thus the resulting function $\psi$ satisfies the three required thermodynamic conditions.

\subsection{Loss function and training}
\label{Sec:Loss}

Analogously to Physics Informed Neural Networks (PINNs), we build the loss function as the sum of the squared residuals of the equations governing the dynamics (PDEs) 
and the boundary conditions (BCs). To be more precise, denoting $\boldsymbol{\xi} \coloneqq \left\{ \mathbf{z}, \mathbf{w} \right\}$ and $\mathbf{p} \coloneqq \left\{f, \psi \right\}$, and defining $\hat{\boldsymbol{\xi}}$ as the combination of $\boldsymbol{\xi}$ and the data for the boundary conditions (such as boundary tractions), 
we abstractly write the PDEs and BCs obtained from Onsager's variational principle as
\begin{equation}
\begin{split}
    & \mathcal{A}_k \mathbf{p} 
    \left( \boldsymbol{\xi} \right) = 0,      
    \quad \quad 
    k = 1, \dots, n_{PDEs}, \quad \text{and} \\
    & \mathcal{B}_k \mathbf{p}
    \left( \boldsymbol{\xi} \right) = h\left(\hat{\boldsymbol{\xi}}\right),
    \quad \quad 
    k = 1, \dots, n_{BCs}.
\end{split}
\end{equation}
Here, $n_{PDEs}$ and $n_{BCs}$ are the numbers of PDEs and BC equations, respectively; $\mathcal{A}_k$ and $\mathcal{B}_k$ represent the operators for the corresponding equations; and $h\left(\hat{\boldsymbol{\xi}}\right)$ contains the boundary data. Further denoting the input data for the PDEs and BCs as $\left\{ \boldsymbol{\xi}_{PDEs}^j \right\}_{j=1}^{N_{PDEs}}$ and $\left\{ \hat{\boldsymbol{\xi}}_{BCs}^j \right\}_{j=1}^{N_{BCs}}$, respectively, the loss function to be minimized may then be expressed as
\begin{equation}
\label{Eq:Loss}
    \mathcal{L} = \boldsymbol \alpha_{PDEs} \cdot \mathcal{L}_{PDEs} + \boldsymbol \alpha_{BC} \cdot \mathcal{L}_{BCs}, 
\end{equation}
where the loss vectors, $\mathcal{L}_{PDEs}$ and $\mathcal{L}_{BCs}$, have components
\begin{equation}
\label{Eq:Loss_PDEs}
    \mathcal{L}_{PDEs}^k 
    = \frac{1}{N_{PDEs}} \sum_{j=1}^{N_{PDEs}} 
    \left| \mathcal{A}_k \mathbf{p} \left(\boldsymbol{\xi}^j_{PDEs} ; \boldsymbol \theta\right) \right|^2 ,
    \quad \quad
    k = 1, \dots, n_{PDEs}, \quad \text{and}
\end{equation}
\begin{equation}
\label{Eq:Loss_BCs}
    \mathcal{L}_{BCs}^k 
    = \frac{1}{N_{BCs}} \sum_{j=1}^{N_{BCs}} 
    \left| \mathcal{B}_{k} \mathbf{p} \left(\boldsymbol{\xi}^j_{BCs}; \boldsymbol \theta\right) 
    - h\left( \hat{\boldsymbol{\xi}}_{BCs}^j \right) \right|^2 ,
    \quad \quad
    k = 1, \dots, n_{BCs},
\end{equation}
and $\boldsymbol \alpha_{PDEs}$ and $\boldsymbol \alpha_{BC}$ denote the corresponding weight vectors for the loss terms. In these equations, the vector $\boldsymbol \theta = \{ \boldsymbol \theta_f, \boldsymbol \theta_{\psi}\}$ encompasses the collection of all the parameters of the the neural networks and $\mathbf{p}(\boldsymbol{\xi}; \boldsymbol{\theta})$ denotes the approximation of $\mathbf{p}(\boldsymbol{\xi})$ by the networks. 


The loss weights $\boldsymbol \alpha_{PDEs}$ and $\boldsymbol \alpha_{BC}$ just introduced are often taken as constants and manually chosen after a time-consuming fine-tuning process. Yet, these are critical to the training, and with an improper choice, PINNs usually fail to achieve stable and accurate results. In order to increase the robustness of the learning process, we use a recently developed method with adaptive loss weights \citep{wang2022and} for each loss term during the training. In this method, the weights are chosen based on the trace (or the summation of the eigenvalues) of the diagonal matrix blocks in the Neural Tangent Kernel (NTK) of the PINNs. For our architecture, the adaptive loss weights are given by
\begin{equation}
\label{Eq:alpha_PDEs}
    \boldsymbol \alpha^k_{PDEs} 
    = \frac{\text{tr} \left( \mathbf{K} \right)}{\text{tr} \left( \mathbf{K}_{kk} \right)},
    \quad \quad
    k = 1, \dots, n_{PDEs},
\end{equation}
\begin{equation}
\label{Eq:alpha_BCs}
    \boldsymbol \alpha^k_{BCs} 
    = \frac{\text{tr} \left( \mathbf{K} \right)}{\text{tr} \left( \mathbf{K}_{ll} \right)} 
    \quad 
    \text{with }
    l=k+n_{PDEs},
    \quad \quad
    k = 1, \dots, n_{BCs},
\end{equation}
where $\text{tr}(\cdot)$ denotes the trace of a matrix, and $\mathbf{K}$ is the NTK matrix that consists of $\left(n_{PDEs} + n_{BCs} \right) \times \left(n_{PDEs} + n_{BCs} \right)$ matrix blocks $\mathbf{K}_{kl}$. Since the calculation of the weights $\boldsymbol \alpha_{PDEs}$ and $\boldsymbol \alpha_{BC}$ only requires the traces for the matrix blocks in the diagonal, we here only provide the formula of the traces for simplicity. These read
\begin{equation}
\label{Eq:tr_NTK_kk}
    \text{tr} \left( \mathbf{K}_{kk} \right) =
    \begin{dcases}
      \sum_{j=1}^{N_{PDEs}} \sum_{\theta \in \boldsymbol \theta}
      \left| \frac{\partial \mathcal{A}_k \mathbf{p}(\boldsymbol{\xi}_{PDEs}^j; \boldsymbol \theta)}{\partial \theta} \right|^2,
      & \text{if } k = 1, \dots, n_{PDEs}, \\
      \sum_{j=1}^{N_{BCs}} \sum_{\theta \in \boldsymbol \theta}
      \left| \frac{\partial  \mathcal{B}_{k-n_{PDEs}} \mathbf{p}(\boldsymbol{\xi}_{BCs}^j; \boldsymbol \theta)}{\partial \theta} \right|^2,
      & \text{if } k = n_{PDEs} + 1, \dots, n_{PDEs} + n_{BCs},
    \end{dcases}  
\end{equation}
\begin{equation}
\label{Eq:tr_NTK}
    \text{tr} \left( \mathbf{K} \right) 
    = \sum_{k=1}^{n_{PDEs}+n_{BCs}} 
    \text{tr} \left( \mathbf{K}_{kk} \right).
\end{equation}
We remark that $h\left( \cdot \right)$ does not participate in the derivatives above, as it was considered to be independent of $\boldsymbol \theta$. Yet, these expressions could be easily generalized in the event of such a dependence.

Equations \eqref{Eq:alpha_PDEs}-\eqref{Eq:tr_NTK} indicate that the loss weights $\boldsymbol \alpha_{PDEs}$ and $\boldsymbol \alpha_{BCs}$ depend on the training parameters $\boldsymbol \theta$ and, hence, these should be updated during the  training process\footnote{We note that these loss weights should be taken as constants when evaluating the gradient of the loss to ensure that the desired optimal value for the network parameters $\mathbf{\theta}$ is obtained. That is $\nabla \mathcal{L} 
    = \boldsymbol \alpha_{PDEs} \cdot 
    \nabla \mathcal{L}_{PDEs} 
    + \boldsymbol \alpha_{BC} \cdot \nabla \mathcal{L}_{BCs}.$}.
However, according to the theory by \cite{wang2022and}, these adaptive loss weights are not expected to change much during the training. In practice, we have observed that although these weights may change by an order of magnitude in some of the following examples, updating such weights only slightly reduces the loss during the training. Therefore, for simplicity, we only compute the adaptive loss weights at the beginning of the training 
and treat them as constants during the training process. To this regard, we note that all the weights in the NNs (INN, PICINN or FICINN) are initialized according to the Glorot initialization scheme and all the biases are initialized as zero.

\subsection{Hardware and implementation}
JAX is the main Python library for neural networks implementation in this paper, while some standard libraries such as Numpy, Scipy, Pandas, and Matplotlib are used for data pre- and post-processing. Since JAX brings together the Autograd and XLA compiler, along with the Just-in-Time (JIT) compilation which comes hand in hand with XLA, it saves numerous computational power and allows us to perform all of the following experiments simply using the NVIDIA TESLA P100 GPU on Google Colab. 

\section{Example 1. Phase transformation of coiled-coil proteins} \label{Sec:PhaseTransformation}

Coiled-coils is a prevalent structural motif occurring in about $10 \%$ of the proteins, where $\alpha$-helices are wrapped around each other \citep{rose2004scaffolds,torres2019combined}. An important mechanical feature, key to several biological functions and emerging biomaterials, is its capability to undergo structural transformations, from the coiled state to an unwinded state. This phase transformation is characterized by a double-well free energy landscape, which may be obtained from molecular dynamic simulations, and it is also strongly affected by hydrodynamic interactions with the solvent. In this example, we will aim at recovering the non-convex free energy density and the dissipation potential densities characterizing the hydrodynamic interactions, directly from synthetic trajectory data in pulling experiments. This could be particularly insightful induced by the large disparities between pulling velocities in non-equilibrium molecular dynamic simulations, and those used in experiments.

\subsection{Model description}
\label{Sec:Model_PhaseTrans}

We here model the mechanical response of the protein as a one-dimensional rod with length $L$, which is fixed at one end, $X=0$, and pulled at the other end, $X=L$, as in \cite{torres2019combined}. In this example, the free energy density can be written as a (nonconvex) function of the strain $\varepsilon$, i.e., $f = f(\varepsilon)$, and the dissipation potential density as a function of the velocity $v$, i.e., $\psi = \psi(v)$. Furthermore, body forces and inertia may be considered negligible. Hence, by Onsager's variational principle, the evolution of the system is governed by
\begin{align}
\label{Eq:Equi_PhaseTrans}
    &\frac{\partial f'(\varepsilon(X,t))}{\partial X}   = \psi'(v(X,t)), \\ \label{Eq:traction_PhaseTrans}
    &    \bar{t} = f'(\varepsilon(L,t)).
\end{align}


\subsection{Data generation}
\label{Sec:DataGeneration_PhaseTrans}

\begin{figure}[H]
    \centering
    \includegraphics[width=\textwidth]{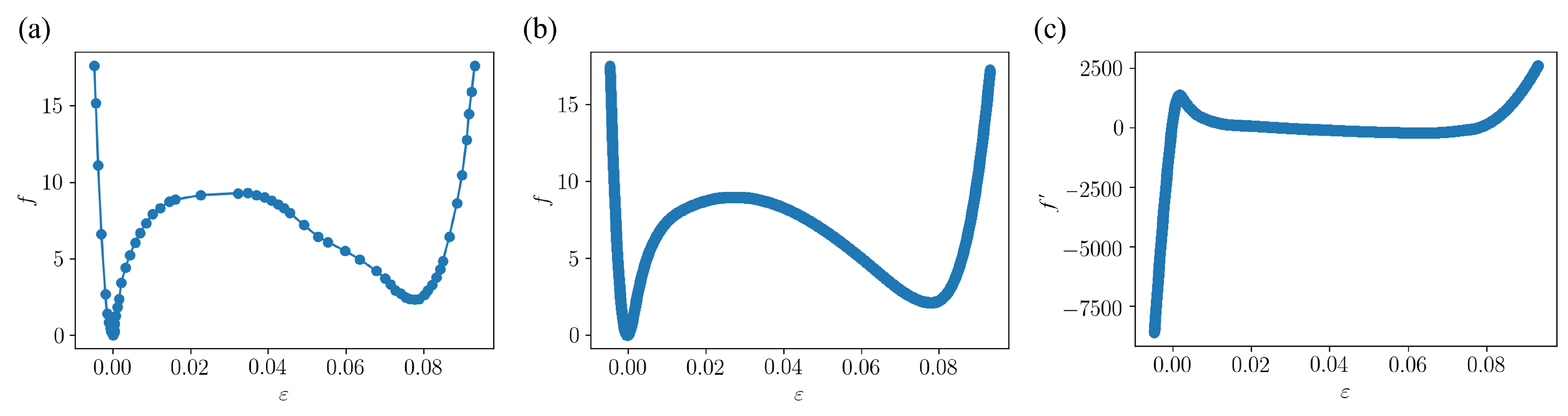}
    \caption{Free energy density of \cite{torres2019combined}, obtained from molecular dynamics simulations. (a) Raw data obtained from the reference. (b) Data at 50,000 equispaced strain values, interpolated from (a). (c) External force, interpolated from (a) at 50,000 equispaced strain values.}
    \label{Fig:protein_ref_data}
\end{figure}

To generate the training data, we use the double-well free energy landscape obtained from molecular dynamic simulations in \cite{torres2019combined}, shown in Fig.~\ref{Fig:protein_ref_data}(a), as well as their estimated dissipation potential $\psi(v)=\frac{1}{2} \eta v^2$, with $\eta=8$ pN ns nm$^{-2}$. The rod size is $L=9$ nm and the pulling velocity at $X=L$ is $v_p/2$, where $v_p=9.5$ m/s. 

To describe the double-well free energy profile in  \cite{torres2019combined}, 61 raw data points are captured (Fig.~\ref{Fig:protein_ref_data}a) from their image. Then, 50,000 equispaced strain values are generated, and the corresponding values for the free energy density (Fig.~\ref{Fig:protein_ref_data}b) and its derivative (Fig.~\ref{Fig:protein_ref_data}c) are obtained by interpolation using the B-spline method (function \texttt{interpolate.splrep} in Scipy, with smoothing condition $s=3$). Finally, the analytic function of the free energy density $f(\varepsilon)$ and its derivative $f'(\varepsilon)$ are defined as the linear interpolation of this finer dataset. The function $f'(\varepsilon)$ is used for data generation, see Eq.~\eqref{Eq:Equi_PhaseTrans}, while $f(\varepsilon)$ is later used for validation purposes.


Equation \eqref{Eq:Equi_PhaseTrans} is solved by the following finite difference scheme
\begin{equation}
\label{Equi_PhaseTrans_discr}
    \frac{f'(\varepsilon_{i+1}^n) - f'(\varepsilon_{i}^n)}{\Delta X}
    = \psi'(v_i^n)
    \quad \text{for} \quad
    i = 1, \dots, N_X - 1
\end{equation}
with 
\begin{equation}
\label{Eq:strain_v_PhaseTrans_discr}
    \varepsilon_i^n = \frac{u_i^n - u_{i-1}^n}{\Delta X},
    \quad \quad
    v_i^n = \frac{u_i^{n+1} - u_{i}^n}{\Delta t}
    \quad \text{for} \quad
    i = 1, \dots, N_X - 1,
\end{equation}
\begin{equation}
    u_0^n = 0
    \quad \text{and} \quad
    u_{N_X}^n = v_p t^n/2,
\end{equation}
where the 1D domain $[0,L]$ is discretized into $N_X+1$ equispaced points ($\Delta X = L/N_X$), and time is discretized uniformly as well with a time step $\Delta t$. Here, $u$ denotes the displacement, and subscripts and superscripts refer to the spatial and temporal indices, respectively, i.e., $u_i^n=u(X_i, t^n)$.

The traction at the boundary may then be computed as
\begin{equation}
\label{Eq:traction_PhaseTrans_discr}
    \bar{t}^n = f'(\varepsilon^n_{N_X}).
\end{equation}

\begin{figure}[H]
    \centering
    \includegraphics[width=0.8\textwidth]{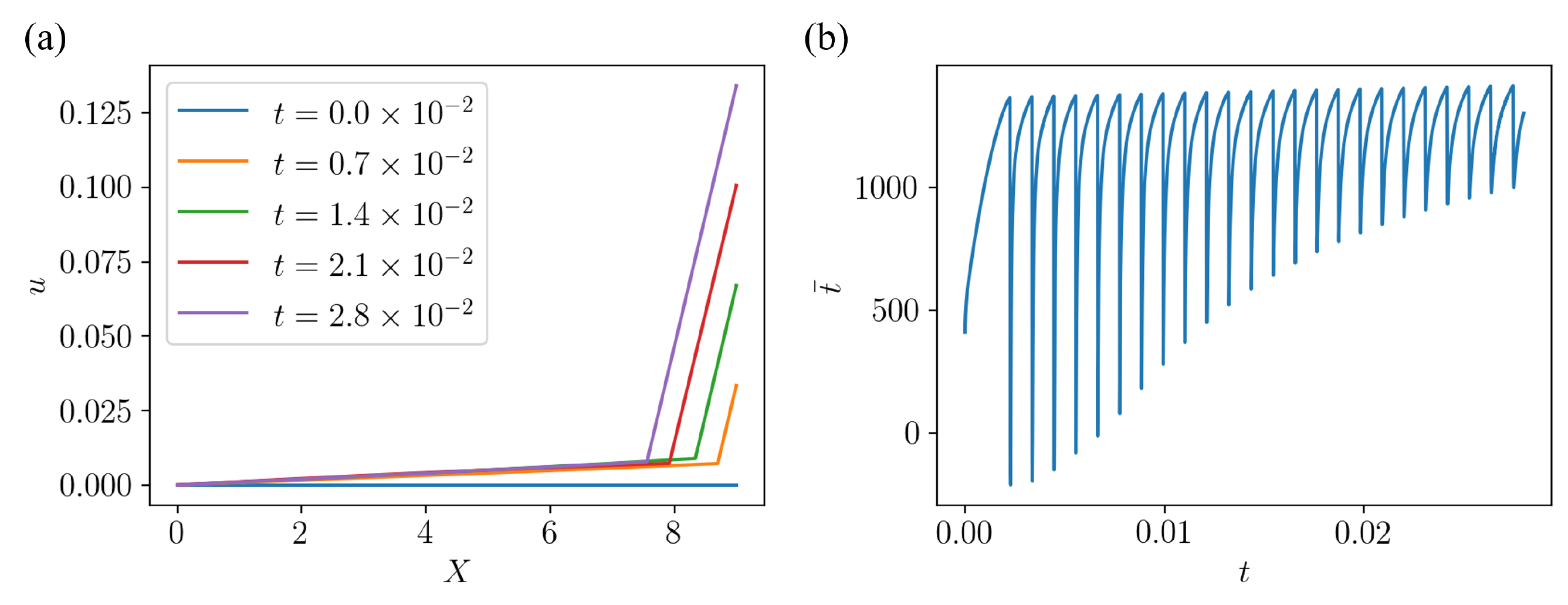}
    \caption{Generated data for the phase transformation model. (a) Snapshots of the displacement field. (b) Applied traction at the pulling end.}
    \label{Fig:u_traction_PhaseTrans}
\end{figure}

In the numerical experiment, we choose $N_X=150$, $\Delta t=9\times 10^{-9}$ ns and take the total simulation time as $T = 0.028$ ns. During the simulation, the displacement field $u_i^n$ and the traction $\bar{t}^n$ at the pulling end are recorded, and the corresponding results are shown in Fig.~\ref{Fig:u_traction_PhaseTrans}. A propagating sharp discontinutiy in the displacement field can be clearly observed, indicative of the phase transformation taking place in the protein.


\subsection{Networks architecture and loss function}
\label{Sec:Architecture_PhaseTrans}

For this phase transformation problem we use an INN to represent the non-dimensional free energy density $\tilde{f}(\tilde{\varepsilon})$ and a FICINN to represent the non-dimensional dissipation potential density $\tilde{\psi}(\tilde{v})$, each with 2 hidden layers and 25 neurons per layer. 
Their dimensional analogues are then computed by means of Eqs.~\eqref{Eq:rescale_fe} and \eqref{Eq:rescale_psi}, where the characteristic scales $f^*$ and $\psi^*$ are calculated from the data on the basis of dimensional analysis as 
\begin{equation}
    f^* = \sigma_{\bar{t}} \sigma_{\varepsilon_{BC}},
\end{equation}
\begin{equation}
    \psi^* = \frac{\sigma_{\bar{t}} \sigma_{v}}{L}.
\end{equation}
Here, $\sigma_{v}$ is the standard deviation of $v_i^n$ used for $\mathcal{L}_{PDE}$, $\sigma_{\varepsilon_{BC}}$ is the standard deviation of $\varepsilon_{N_X}^n$ used for $\mathcal{L}_{BC}$ and $\sigma_{\bar{t}}$ is the standard deviation of the input data of $\bar{t}^n$ on the boundary. 

The loss function $\mathcal{L}$ is defined based on the residual of the discretized Eqs.~\eqref{Equi_PhaseTrans_discr} and \eqref{Eq:traction_PhaseTrans_discr} as,
\begin{equation}
\label{Eq:Loss_PhaseTrans}
    \mathcal{L} = \alpha_{PDE} \mathcal{L}_{PDE} 
    + \alpha_{BC} \mathcal{L}_{BC},
\end{equation}
with 
\begin{equation}
\label{Eq:Loss_PDE_PhaseTrans}
    \mathcal{L}_{PDE} 
    = \frac{1}{N_{PDE}} \sum_{j=1}^{N_{PDE}}  
    \left| \frac{f'(\varepsilon_{i+1}^n; \boldsymbol \theta_f) - f'(\varepsilon_{i}^n; \boldsymbol \theta_f)}{\Delta X}
    - \psi'(v_i^n; \boldsymbol \theta_{\psi}) \right|^2,
\end{equation}
\begin{equation}
\label{Eq:Loss_BC_PhaseTrans}
    \mathcal{L}_{BC}
    = \frac{1}{N_{BC}} \sum_{j=1}^{N_{BC}} 
    \left| \bar{t}^n - f'(\varepsilon^n_{N_X}; \boldsymbol \theta_f) \right|^2.
\end{equation}
Here, we have used the notation introduced in Sec.~\ref{Sec:Loss}, and defined the input dataset for the PDE, $\{\boldsymbol{\xi}_{PDE}^j\}_{j=1}^{N_{PDE}}$ as $\boldsymbol{\xi}_{PDE}^j = \left( \varepsilon_i^n, \varepsilon_{i+1}^n, v_i^n \right)$, with $N_{PDE} \leq (N_X-1)n_T$ and $j$ a one-dimensional index that denotes the flattened order of a subset of the two-dimensional index $(i,n)$, with $i=1, \dots, N_X-1$ and $n = 0, \dots, n_T - 1$. Similarly,  the input dataset for the boundary conditions, $\{\hat{\boldsymbol{\xi}}_{BC}^j\}_{j=1}^{N_{BC}}$, is defined as $\hat{\boldsymbol{\xi}}_{BC}^j = \left( \varepsilon_{N_X}^n, \bar{t}^n \right)$ with $N_{BC}\leq n_T$, and $j$ an index over a subset of $n=0, \dots, n_T-1$. The following section will describe in detail the choice of the dataset used for the training process, aimed at reducing the sampling bias \citep{mehrabi2021survey}. Finally, we remark that the neighbor information $\varepsilon_{i+1}^n$ is packed together with the local data at $(i, n)$ in $\boldsymbol{\xi}_{PDE}^j$. Although this requires additional space to save the data, this strategy largely reduces the computational cost when calculating the spatial gradient. 


\subsection{Data pre-processing strategy}
\label{Sec:PreProcessing_PhaseTrans}

\begin{figure}[H]
    \centering
    \includegraphics[width=0.4\textwidth]{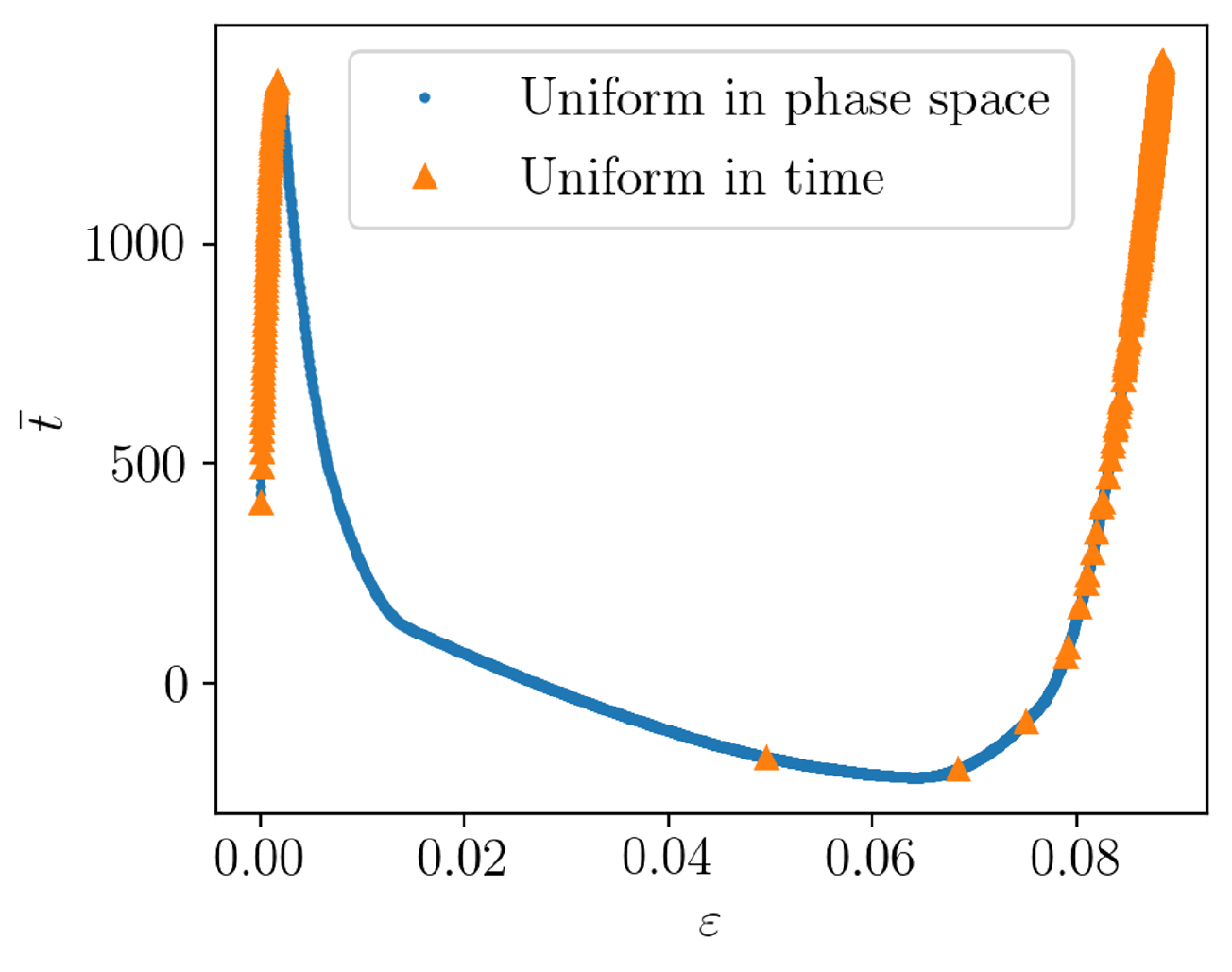}
    \caption{Comparison of the BC data when selected uniformly in phase-space (blue dots) versus uniformly in time (orange triangles).}
    \label{Fig:protein_data_compare}
\end{figure}

Despite the apparent simplicity of this one-dimensional problem, the output dataset nears $470$ million data entries, induced by the small time step required for numerical stability. This makes it nearly impossible, or at least impractical, to train with the whole output dataset. A na{\"i}ve approach to this problem is to select the data uniformly with a larger time step. However, this approach fails to predict the correct free energy density and dissipation potential density, as shown in  Fig.~\ref{Fig:equispaced_protein_prediction} of Appendix \ref{App:PhaseTrans_Coarser_t_Constant_Weights}. The underlying reason is that the double-well nature of the free energy landscape leads to highly biased data, as depicted in Fig.~\ref{Fig:protein_data_compare}.


To address the above issue, we here implement a data pre-processing strategy that selects data uniformly in the input space of the functions to be trained. According to Eqs.~\eqref{Eq:Loss_PhaseTrans}-\eqref{Eq:Loss_BC_PhaseTrans}, the free energy density $f(\varepsilon)$ is primarily determined from the BC, while the dissipation potential density $\psi(v)$ is learned from the PDE. Therefore, for the boundary dataset $\{\hat{\boldsymbol{\xi}}_{BC}^j\}_{j=1}^{N_{BC}}$, we select $N_{BC}$ boundary data $\hat{\boldsymbol{\xi}}_{BC}^j$ such that the corresponding boundary strains $\varepsilon_{N_X}^n$ are distributed uniformly within its range. Similarly, for the PDE dataset $\{\boldsymbol{\xi}_{PDE}^j\}_{j=1}^{N_{PDE}}$, we select $N_{PDE}$ data $\boldsymbol{\xi}_{PDE}^j$ such that the corresponding velocities $v_n^i$ are distributed uniformly. In both cases, this uniform data selection is achieved by the following steps: (i) generate a uniform mesh grid with $N_{BC}$ (or $N_{PDE}$) nodes spanning the full range of boundary strain $\varepsilon_{N_X}^n$ (or velocity $v_n^i$), (ii) search the closest boundary strain $\varepsilon_{N_X}^n$ (or velocity $v_n^i$) for each node, and (iii) save the indices of the selected data (eliminating duplicates) and record the boundary data $\hat{\boldsymbol{\xi}}_{BC}^j$ (or PDE data $\boldsymbol{\xi}_{PDE}^j$) corresponding to those indices. 

\begin{figure}[H]
    \centering
    \includegraphics[width=0.8\textwidth]{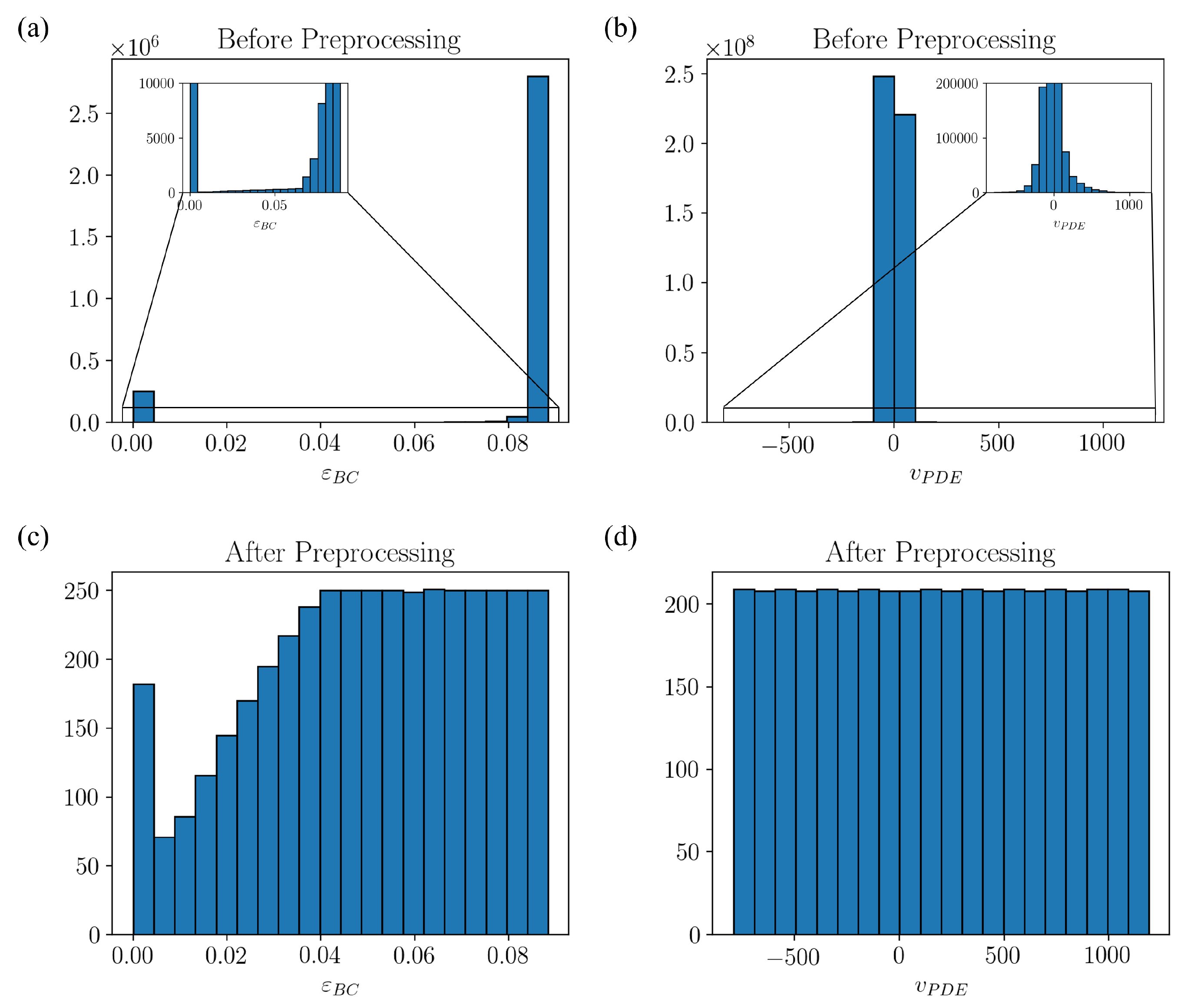}
    \caption{Data for the pulling experiment of a protein undergoing a phase transformation before the pre-processing strategy ((a) and (b)) and after the pre-processing strategy ((c) and (d)).}
    \label{Fig:protein_training_data}
\end{figure}

In this model we choose $4170$ for both boundary and PDE data, 80$\%$ of which (randomly selected) is used for training purposes ($N_{BC} = N_{PDE} = 3336$), while the remaining 20$\%$ is used  for testing to prevent the model from overfitting. Figure \ref{Fig:protein_training_data} shows a comparison of the data histograms before and after the pre-processing strategy, clearly showing a reduction in the sampling bias. This may be also observed in Fig.~\ref{Fig:protein_data_compare} for the boundary traction data as a function of the boundary strain data, where the data using the pre-processing strategy and that obtained from a uniform larger time step are compared (see Appendix \ref{App:PhaseTrans_Coarser_t_Constant_Weights} for further discussions).


\subsection{Training and results}
\label{Sec:Result_PhaseTrans}

For this model, we train both the INN and FICINN for 30,000 epochs using an Adam optimizer with learning rate of $10^{-4}$. 
\begin{figure}[h]
    \centering
    \includegraphics[width=0.8\textwidth]{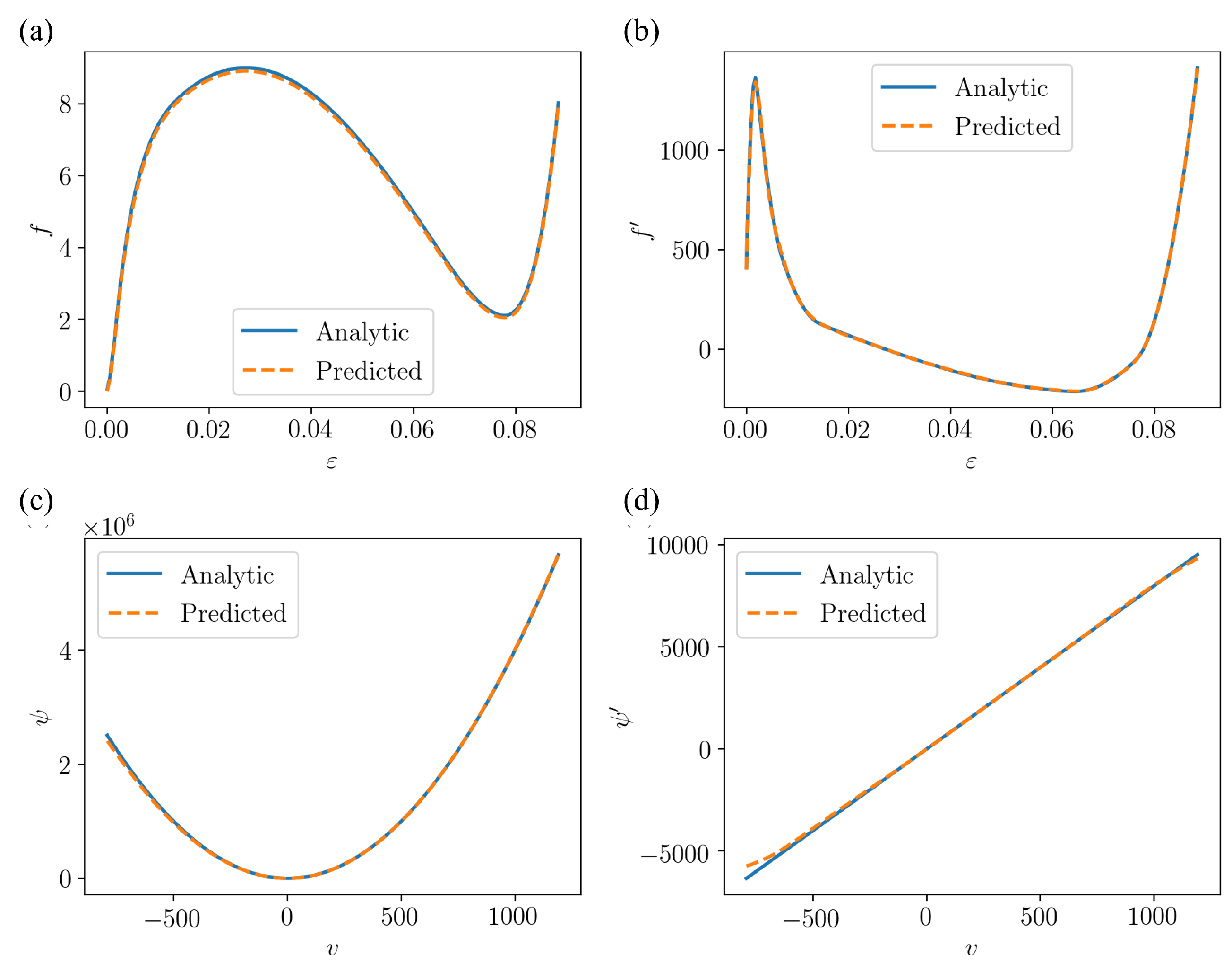}
    \caption{Comparison between analytic values and predictions from the training model for (a) free energy density $f(\varepsilon)$, (b) stress $f'(\varepsilon)$, (c) dissipation potential density $\psi(v)$ and (d) viscous force $\psi'(v)$.}
    \label{Fig:protein_prediction}
\end{figure}
The training results are shown in Fig.~\ref{Fig:protein_prediction}, where the predictions are shown to have a very good agreement with the analytic values for the free energy density $f(\varepsilon)$, the dissipation potential density $\psi(v)$ and their derivatives. Indeed, the relative $L^2$ errors for $f(\varepsilon)$, $f'(\varepsilon)$, $\psi(v)$ and $\psi'(v)$ within the data range shown are $1.25\%$, $1.55\%$, $1.08\%$ and $2.47\%$, respectively, where this error for a general function $A(x)$ within an interval $[x_1, x_2]$ is defined as
\begin{equation}
    \text{err}_A 
    = \frac{\int_{x_1}^{x_2} \left|A_{ana}(x) - A_{pred}(x)\right|^2 dx}{\int_{x_1}^{x_2} \left|A_{ana}(x)\right|^2 dx}
    \times 100\%.
\end{equation} 

To further emphasize the importance of the adaptive loss weights on the training process, we show the results of the training with constant loss weights, guessed on the basis of dimensional analysis, in Fig.~\ref{Fig:static_protein_prediction} of Appendix \ref{App:PhaseTrans_Coarser_t_Constant_Weights}. In this case, the analytic functions for $f(\varepsilon)$ and $\psi(v)$ cannot be recovered, and it is also found that the results become very sensitive to the choice of the loss weights.


\section{Example 2. Dynamic response of a viscoelastic rod}
\label{Sec:Viscoelastic}

Viscoelasticity is one of the most ubiquitous inelastic phenomena in materials, yet, it is simultaneously, the one whose models encode some of the strongest phenomenological assumptions. Such models, or parameters within, are thus most often identified from experimental observations, which makes it a task that is particularly well suited for machine learning techniques. In this example, we aim at recovering the free energy and dissipation potential densities from synthetic data of the dynamic response of a three-dimensional viscoelastic rod; in particular, we aim at obtaining a reduced one-dimensional model. To this regard we note that Onsager's variational principle may be used as an approximation tool to obtain an approximate description of the evolution equations when a reduced representation of the system is used \citep{doi2015onsager}.


\subsection{Model description}
\label{Sec:Viscoelastic_Model}

We consider a three-dimensional viscoelastic material, whose elastic response is described by a linear isotropic material with bulk and shear modulus $K$ and $G$, respectively, while the viscoelastic response is modeled by means of a Prony series with viscous shear moduli $G_{v\alpha}$ and relaxation times $\tau_{\alpha}$ for $\alpha=1, \dots, n$. Its constitutive relation may then be written as 
\begin{equation}
\label{Eq:StressStrain}
    \boldsymbol{\sigma} = 3 K \pmb{\varepsilon}^{vol} + 2 G \pmb{\varepsilon}^{dev} 
    + \sum_{\alpha=1}^n 2 G_{v\alpha} \left( \pmb{\varepsilon}^{dev} - \pmb{\varepsilon}^{v \alpha} \right),
\end{equation}
where $\boldsymbol \sigma$ is the stress tensor, $\pmb \varepsilon^{vol} \coloneqq \frac{1}{3} \text{tr}(\pmb{\varepsilon}) \mathbf{I}$ and $\pmb \varepsilon^{dev} \coloneqq \pmb{\varepsilon} - \pmb{\varepsilon}^{vol}$ are the volumetric and deviatoric strain tensors, respectively, and the internal variables (viscous strains $\pmb{\varepsilon}^{v \alpha}$) obey the following evolution equations
\begin{equation}
\label{Eq:ViscousStrainRelaxation}
    \tau_{\alpha} \dot{\pmb{\varepsilon}}^{v \alpha}
    = \pmb{\varepsilon}^{dev} - \pmb{\varepsilon}^{v \alpha}, 
    \quad \quad 
    \alpha = 1, \dots, n.
\end{equation}


The equilibrium equations and evolution equations \eqref{Eq:ViscousStrainRelaxation} have a variational characterization {\`a} la Onsager, with state and process variables $\mathbf{z} = \left\{ \pmb{\varepsilon}, \pmb{\varepsilon}^{v1}, \dots, \pmb{\varepsilon}^{vn} \right\}$ and  $\mathbf{w} = \left\{ \mathbf{v}, \dot{\pmb{\varepsilon}}^{v1}, \dots, \dot{\pmb{\varepsilon}}^{vn} \right\}$, and with the following free energy and dissipation potential densities
\begin{equation}
    f_{3D} \left( \pmb{\varepsilon}, \{\pmb{\varepsilon}^{v \alpha}\}_{\alpha=1}^n \right)
    = \frac{1}{2} K \left( \text{tr} \pmb{\varepsilon} \right)^2
    + G \pmb{\varepsilon}^{dev} : \pmb{\varepsilon}^{dev}
    + \sum_{\alpha=1}^n G_{v\alpha}
    \left( \pmb{\varepsilon}^{dev} - \pmb{\varepsilon}^{v \alpha} \right) 
    : \left( \pmb{\varepsilon}^{dev} - \pmb{\varepsilon}^{v \alpha} \right) 
\end{equation}
\begin{equation}
    \psi_{3D} \left( \{\dot{\pmb{\varepsilon}}^{v\alpha}\}_{\alpha=1}^n \right)
    = \sum_{\alpha=1}^n \frac{1}{2} \eta_{\alpha}
    \dot{\pmb{\varepsilon}}^{v\alpha}
    : \dot{\pmb{\varepsilon}}^{v\alpha}, \quad \text{where $\eta_\alpha = 2G_{v\alpha} \tau_\alpha$}.
\end{equation}
That is, these evolution equations may be cast as
\begin{equation}
\label{Eq:Viscoelastic_PDE_dynamics}
    \nabla \cdot \boldsymbol \sigma 
    = \nabla \cdot 
    \left(  \frac{\partial f_{3D}}{\partial \pmb{\varepsilon}} \right)
    = \rho \textbf{a} 
\end{equation}
\begin{equation}
\label{Eq:Viscoelastic_PDE_constitutive}
    \frac{\partial \psi_{3D}}{\partial \dot{\pmb{\varepsilon}}^{v\alpha}}
    + \frac{\partial f_{3D}}{\partial \pmb{\varepsilon}^{v\alpha}}  = 0,
    \quad
    \text{for } \alpha = 1, \dots, n,
\end{equation}
where $\textbf{a}$ is the acceleration.

The specific example here considered consists of a slender three-dimensional viscoelastic bar of the type just described, with one end fixed to a wall (only normal displacements to the wall are zero), while the other end is subjected to a prescribed excitation along the direction of the bar (defined as direction 1). This problem naturally admits a one-dimensional characterization, that we will aim at discovering. For validation purposes, we have analytically derived in Appendix \ref{App:3D_to_1D} the one-dimensional constitutive equation and evolution equations. Denoting for simplicity $\sigma = \sigma_{11}$, $\varepsilon = \varepsilon_{11}$ and $\varepsilon^{v \alpha} = \varepsilon^{v \alpha}_{11}$ for $\alpha = 1, \dots, n$, these equations read
\begin{equation}
\label{Eq:StressStrain_1D}
    \sigma = E_{1D} \varepsilon 
    + \sum_{\alpha=1}^n E_{1D\alpha} \left( \varepsilon - \varepsilon^{v \alpha} \right),
\end{equation}
\begin{equation}
\label{Eq:ViscousStrainRelaxation_1D}
    \tau_{\alpha} \dot{\varepsilon}^{v \alpha} 
    = \varepsilon - \frac{\sigma}{9K}
    - \varepsilon^{v \alpha},
\end{equation}
with
\begin{equation}
\begin{gathered}
    \theta = 1+ \frac{G+\sum_{\alpha=1}^n G_{v\alpha}}{3K},
    \quad 
    E_{1D} = \frac{3G}{\theta}\\
    \text{and} \quad 
    E_{1D\alpha} = \frac{3G_{v\alpha}}{\theta},
    \quad
    \eta_{1D\alpha} = \frac{3}{2}\eta_{\alpha}
    \quad 
    \text{for } \alpha = 1, \dots, n.
\end{gathered}
\end{equation}
Equation \eqref{Eq:ViscousStrainRelaxation_1D} and the equilibrium equation may be equivalently written as a function of the one-dimensional free energy and dissipation potential density as
\begin{equation}
\label{Eq:Viscoelastic_PDE_dynamics_1D}
    \frac{\partial}{\partial X}
    \left(  \frac{\partial f_{1D}}{\partial \varepsilon} \right) = \rho a
\end{equation}
\begin{equation}
\label{Eq:Viscoelastic_PDE_constitutive_1D}
    \frac{\partial \psi_{1D}}{\partial \dot{\varepsilon}^{v\alpha}}
    + \frac{\partial f_{1D}}{\partial \varepsilon^{v\alpha}} = 0,
    \quad
    \text{for } \alpha = 1, \dots, n,
\end{equation}
where these potentials read
\begin{equation}
\begin{split}
     f_{1D} \left( \varepsilon, \{\varepsilon^{v\alpha}\}_{\alpha=1}^n \right) &\coloneqq f_{3D}\left( \pmb{\varepsilon}\left(\varepsilon, 
    \left\{ \varepsilon^{v\alpha} \right\}_{\alpha=1}^n \right), \left\{\pmb{\varepsilon}^{v\alpha} \left( \varepsilon^{v\alpha} \right) \right\}_{\alpha=1}^n \right) \\
    & = \frac{1}{2} \theta E_{1D} \varepsilon^2
    + \sum_{\alpha=1}^n \frac{1}{2} \theta E_{1D\alpha} \left( \varepsilon - \varepsilon^{v \alpha} \right)^2
    -  \frac{\theta }{18 K} 
    \left[ E_{1D} \varepsilon + \sum_{\alpha=1}^n E_{1D\alpha} \left( \varepsilon - \varepsilon^{v \alpha} \right) \right]^2 \\
   \psi_{1D} (\{\dot{\varepsilon}^{v\alpha}\}_{\alpha=1}^n) &\coloneqq \psi_{3D} \left( \{ \dot{\pmb{\varepsilon}}^{v\alpha} ((\dot{\varepsilon}^{v\alpha})) \}_{\alpha=1}^n \right)
    = \sum_{\alpha=1}^n \frac{1}{2} \eta_{1D\alpha} \left( \dot{\varepsilon}^{v\alpha} \right)^2.
    \end{split}
\end{equation}
Similarly, the boundary condition can be written as,
\begin{equation}
\label{Eq:Viscoelastic_BC_1D}
    \left. \frac{\partial f_{1D}}{\partial \varepsilon} \right|_{boundary} 
    = \bar{t}.
\end{equation}
That is, the one-dimensional approximate evolution equations, Eqs.~\eqref{Eq:Viscoelastic_PDE_dynamics_1D}, \eqref{Eq:Viscoelastic_PDE_constitutive_1D} and \eqref{Eq:Viscoelastic_BC_1D}, may also be directly recovered from Onsager's variational principle, by selecting the state and process variables as $\mathbf{z} = \left\{ \varepsilon, \varepsilon^{v1}, \dots, \varepsilon^{vn} \right\}$ and $\mathbf{w} = \left\{ v, \dot{\varepsilon}^{v1}, \dots, \dot{\varepsilon}^{vn} \right\}$, respectively. In the following, we will denote the 1D forms of free energy density $f_{1D}$ and dissipation potential density $\psi_{1D}$ as $f$ and $\psi$ so as to simplify the notation and be consistent with the notation used in the network architecture.


\subsection{Data generation and pre-processing}

\begin{figure}[H]
    \centering
    \includegraphics[width=0.7\textwidth]{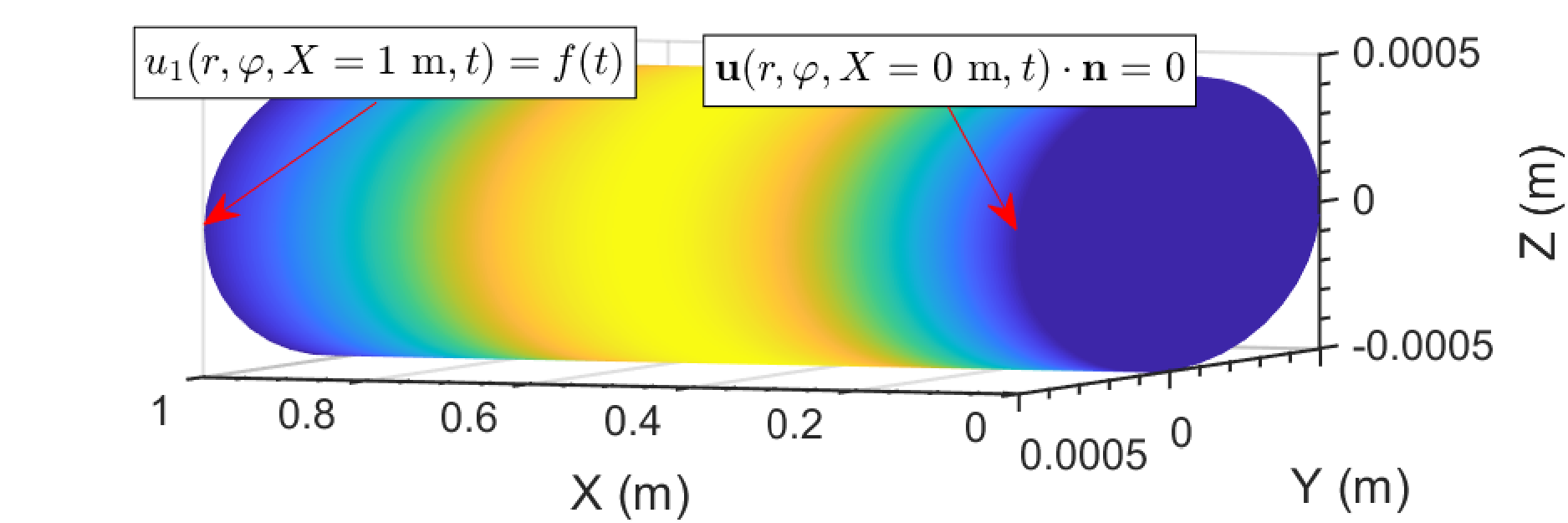}
    \caption{Computational domain used in the simulation of the pulling experiment of a viscoelastic rod.}
    \label{Fig:comp_domain}
\end{figure}

\begin{figure}[H]
    \centering
    \includegraphics[width=0.8\textwidth]{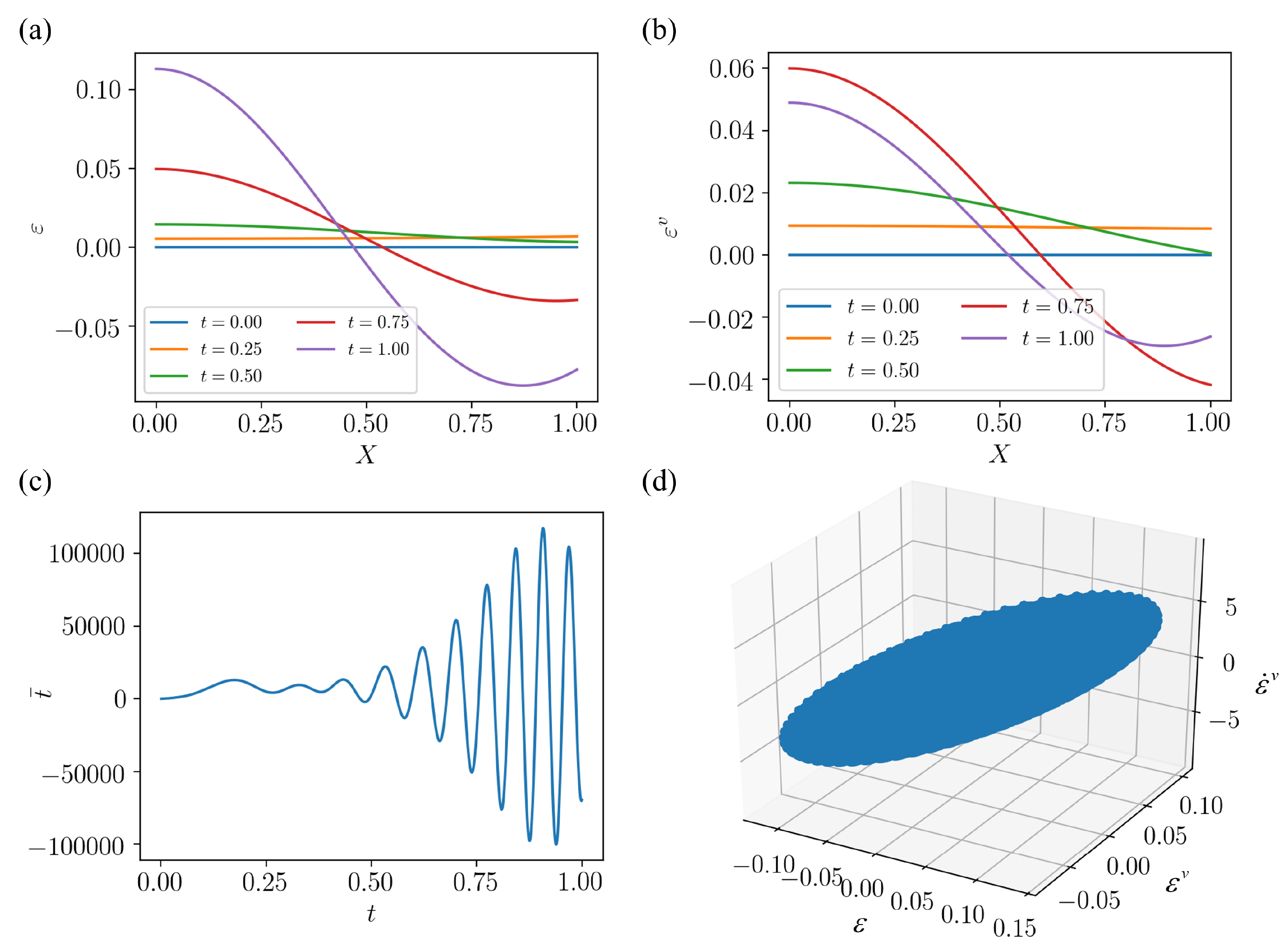}
    \caption{Data for the viscoelastic model. Snapshots for (a) the strain field and (b) the viscous strain field. (c) Time evolution of the applied traction. (d) Trajectories for all spatial points in the phase-space of $\varepsilon$-$\varepsilon^v$-$\dot{\varepsilon}^v$.}
    \label{Fig:viscoelastic_data}
\end{figure}

The pulling experiment previously described is numerically simulated using the finite element method with COMSOL Multiphysics\textsuperscript{\textregistered} \citep{COMSOL54} and the Solid Mechanics Module. 
The computational domain is a bar with diameter 0.001 m and length 1 m, as shown in Fig.~\ref{Fig:comp_domain}, though axial symmetry is used to perform two-dimensional simulations. 
The bar is homogeneous and its constitutive response is given by a generalized Maxwell model with one Maxwell element \citep{christensen2012theory}. The material properties for the cylinder are given in Table \ref{tab:matprops}, and the corresponding bulk and shear moduli can be found as $K=E/\left[ 3 \left( 1-2\nu\right) \right] = 1.25 \times 10^7$~Pa and $G=E/\left[ 2 \left( 1+\nu\right) \right] = 2.52 \times 10^5$~Pa. The mesh contains 750 quadrilateral elements, and quadratic shape functions are used for the finite element analysis. The system starts at rest with a roller boundary condition applied at one end of the cylinder 
and a displacement boundary condition applied to the opposite face. This displacement boundary condition is $u_1(r,\varphi,X=1\textnormal{ m},t)=f(t)=(0.01\textnormal{ m})\left[1-\cos{(2\pi\left( (1 \textnormal{ Hz})+(9\textnormal{ Hz/s})t\right)t)}\right]$ such that the frequency varies from \mbox{1 Hz} to \mbox{10 Hz} in the total \mbox{1 s} of time simulated. The remaining boundaries are subject to free boundary conditions, where $r$, $\varphi$ and $X$ are the radial distance, azimuth and axial coordinate, respectively, and $u_1$ refers to the displacement in the axial direction. A generalized-$\alpha$ method with $\alpha=0.95$ and a time step of \mbox{0.0001 s} is used for the time integration \citep{chung1993}. Data from the simulations is further post-processed by averaging all fields across 251 equally-spaced planes (i.e, $\Delta X =0.004$) perpendicular to the axis of the cylinder at 1000 time steps (i.e, $\Delta t =0.001$). The specific data outputted is strain, viscous strain, acceleration, and the traction at the pulling end, while the viscous strain rate is computed from this data by means of a finite difference scheme. 
These averaged data are then used to obtain the one-dimensional model. The corresponding results for strain $\varepsilon$, viscous strain $\varepsilon^v$, applied traction $\bar{t}$ and the trajectories of each spacial points in the phase space of $\varepsilon$-$\varepsilon^v$-$\dot{\varepsilon}^v$ are shown in Fig.~\ref{Fig:viscoelastic_data}. This data  is randomly split into training data (80\%) and testing data (20\%).

\begin{table}[]
\centering
\begin{tabular}{c|c|c|c|c}
$E$ (Pa)        & $\nu$ & $\rho$ (kg/m$^3$) & $G_v$ (Pa)        & $\tau$ (s) \\ 
\hline
$7.5\times10^5$ & 0.49  & 970               & $7.5\times10^4$ & 0.01      
\end{tabular}
\caption{Density $\rho$, elastic material properties (Young's modulus $E$ and Poisson's ratio $\nu$), and viscoelastic properties (viscous shear modulus $G_v$ and relaxation time $\tau$) used in the pulling experiment.}
\label{tab:matprops}
\end{table}

\subsection{Networks architecture and loss function}
\label{Sec:Architecture_Viscoelastic}

For this viscoelastic problem, we use an INN to represent the non-dimensional free energy density $\tilde{f}(\tilde{\varepsilon}, \tilde{\varepsilon}^v)$ and a FICINN to represent the non-dimensional dissipation potential density $\tilde{\psi}(\dot{\tilde{\varepsilon}}^v)$, each with 2 hidden layers and 25 neurons in each layer. Their dimensional analogues are then computed by means of  Eqs.~\eqref{Eq:rescale_fe_viscoelastic} and \eqref{Eq:rescale_psi}, where the characteristic $f^*$ and $\psi^*$ are calculated from the data on the basis of dimensional analysis as,
\begin{equation}
    f^* = \text{MaxMin}\left(\bar{t}\right)\  \text{MaxMin}\left(\varepsilon_{BC}\right),
\end{equation}
\begin{equation}
    \psi^* = \frac{f^*\ \text{MaxMin}\left( \varepsilon^v_{PDEs}\right)}{\text{MaxMin}\left( \dot{\varepsilon}^v_{PDEs}\right)}.
\end{equation}
Here, $\text{MaxMin}(A)$ refers to the difference between the maximum and the minimum of $A$ within the data, and the subscripts PDEs and BC refer to the data used for the PDEs (related to the interior spacial points, i.e., $i=1, \dots, N_X-1$) and for the BC ($i=N_X$), respectively. 


The loss function $\mathcal{L}$ is defined based on the discretized version of the equilibrium equation \eqref{Eq:Viscoelastic_PDE_dynamics_1D}, the evolution equation for the internal variable  \eqref{Eq:Viscoelastic_PDE_constitutive_1D} and the boundary  equation \eqref{Eq:Viscoelastic_BC_1D} as,
\begin{equation}
\label{Eq:Loss_viscoelastic}
    \mathcal{L} = \alpha_{eq} \mathcal{L}_{eq} 
    + \alpha_{int} \mathcal{L}_{int} 
    + \alpha_{BC} \mathcal{L}_{BC},
\end{equation}
with 
\begin{equation}
\label{Eq:Loss_dyn_viscoelastic}
    \mathcal{L}_{eq} 
    = \frac{1}{N_{PDEs}} 
    \sum_{j=1}^{N_{PDEs}} 
    \left| \frac{f_{,\varepsilon}\left(\varepsilon_{i+1}^n, \left( \varepsilon^v \right)_{i+1}^n; \boldsymbol \theta_f\right) 
    - f_{,\varepsilon}\left(\varepsilon_{i}^n, \left( \varepsilon^v \right)_{i}^n; \boldsymbol \theta_f\right)}{\Delta X}
    - \rho a_i^n  \right|^2,
\end{equation}
\begin{equation}
\label{Eq:Loss_cons_viscoelastic}
    \mathcal{L}_{int} 
    = \frac{1}{N_{PDEs}} 
    \sum_{j=1}^{N_{PDEs}}  
    \left| \psi_{,\dot{\varepsilon}^{v}} \left( \left( \dot{\varepsilon}^v \right)_{i}^n; \boldsymbol \theta_{\psi} \right)
    + f_{,\varepsilon^{v}} \left(\varepsilon_{i}^n, \left( \varepsilon^v \right)_{i}^n; \boldsymbol \theta_f\right)  \right|^2,
\end{equation}
\begin{equation}
\label{Eq:Loss_BC_viscoelastic}
    \mathcal{L}_{BC}
    = \frac{1}{N_{BC}} \sum_{j=1}^{N_{BC}} 
    \left| \bar{t}^n - f_{,\varepsilon} \left(\varepsilon^n_{N_X}, \left( \varepsilon^v \right)^n_{N_X}; \boldsymbol \theta_f \right) \right|^2,
\end{equation}
and
\begin{equation}
    \left(\dot{\varepsilon}^{v}\right)^n_i =  \frac{ (\varepsilon^v)_i^{n+1} -  (\varepsilon^v)_i^n }{ \Delta t}.
\end{equation}
Here, we have used as well the notation introduced in Sec.~\ref{Sec:Loss}, and defined the input dataset for the PDEs, $\{\boldsymbol{\xi}_{PDEs}^j\}_{j=1}^{N_{PDEs}}$, as $\boldsymbol{\xi}_{PDEs}^j = \left( \varepsilon_i^n, \varepsilon_{i+1}^n, \left( \varepsilon^v \right)_{i}^n,  \left( \varepsilon^v \right)_{i+1}^n, \left( \dot{\varepsilon}^v \right)_{i}^n, a_i^n \right)$ with $N_{PDEs} \leq (N_X-1)n_T$ and $j$ as a 1D index that denotes the flattened order of a subset of the 2D index $(i,n)$ with $i=1, \dots, N_X-1$ and $n = 0, \dots, n_T - 1$. Similarly, the input dataset for the boundary conditions, $\{\hat{\boldsymbol{\xi}}_{BC}^j\}_{j=1}^{N_{BC}}$, is defined as $\hat{\boldsymbol{\xi}}_{BC}^j = \left( \varepsilon_{N_X}^n, \left( \varepsilon^v \right)_{N_X}^n, \bar{t}^n \right)$ with $N_{BC}\leq n_T$, and $j$ and index over a subset of $n=0, \dots, n_T-1$. 


\subsection{Training and results}
\label{Sec:Result_Viscoelastic}

\begin{figure}[h]
    \begin{subfigure}
        \centering
        \includegraphics[width=\textwidth]{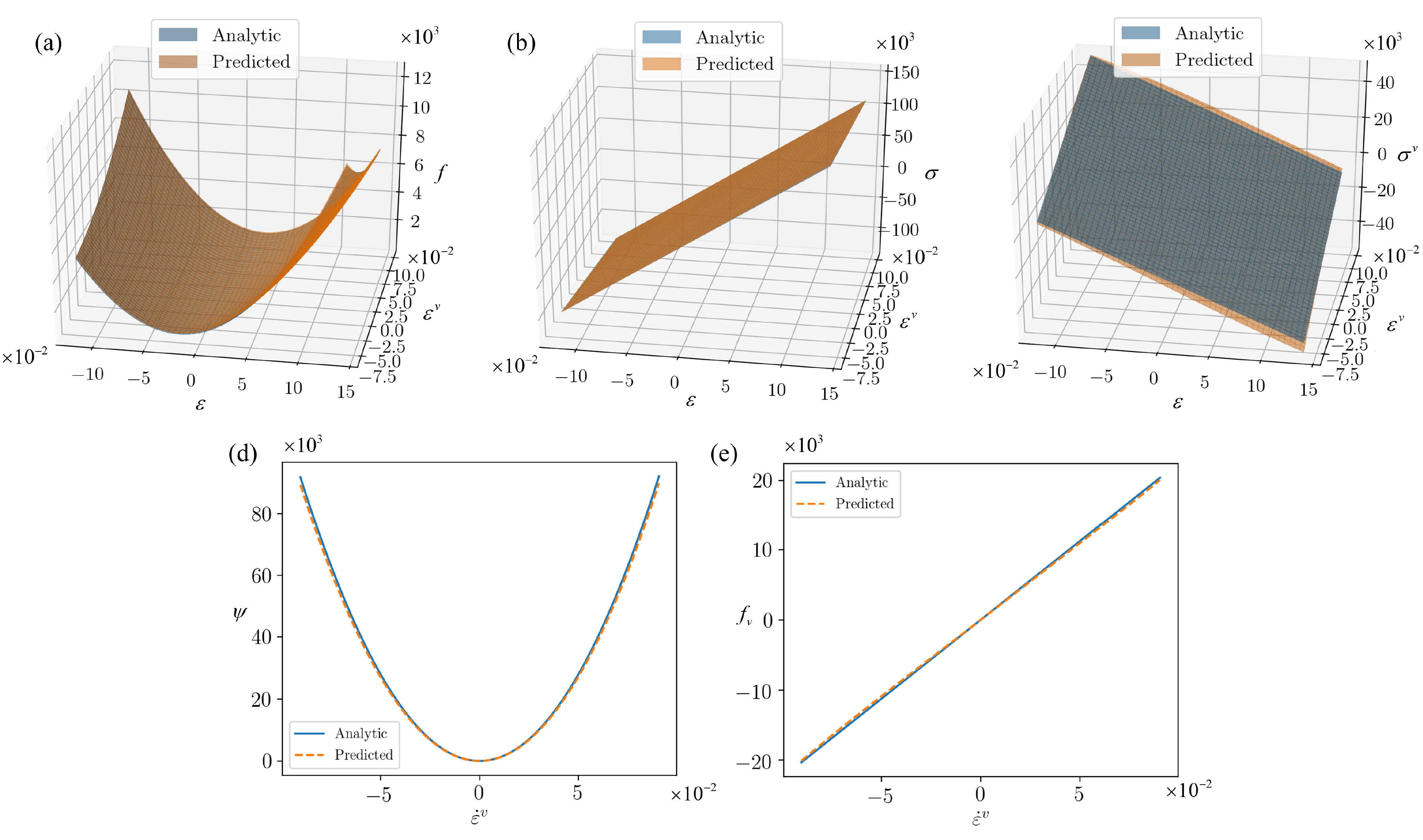}
    \end{subfigure}

    \caption{Comparison between analytic values and predictions from the training model for (a) free energy density $f(\varepsilon, \varepsilon^v)$, (b) stress $\sigma(\varepsilon, \varepsilon^v) = f_{,\varepsilon}(\varepsilon, \varepsilon^v)$, (c) configurational stress $\sigma^v(\varepsilon, \varepsilon^v) = f_{,\varepsilon^v}(\varepsilon, \varepsilon^v)$, (d) dissipation potential density $\psi(\dot{\varepsilon}^v)$ and (e) dissipative force $f_v(\dot{\varepsilon}^v) = \psi'(\dot{\varepsilon}^v)$.}
    \label{Fig:viscoelastic_results}
\end{figure}

For this viscoelastic problem, no data pre-processing is applied, and the neural networks are trained for $30,000$ epochs using an Adam optimizer with a learning rate of $10^{-4}$. Figure~\ref{Fig:viscoelastic_results} shows the training results for the free energy density $f(\varepsilon, \varepsilon^v)$, the dissipation potential density $\psi(\dot{\varepsilon}^v)$ and their derivatives $\sigma(\varepsilon, \varepsilon^v) = f_{,\varepsilon}(\varepsilon, \varepsilon^v)$,  $\sigma^v(\varepsilon, \varepsilon^v) = f_{,\varepsilon^v}(\varepsilon, \varepsilon^v)$, and $f_v(\dot{\varepsilon}^v) = \psi'(\dot{\varepsilon}^v)$. Here $\psi$ and its derivative are plotted within the maximum and minimum value of the training data for $\dot{\varepsilon}^v$, while $f$ and $\sigma$ are plotted within a rectangular domain that bounds the quasi-elliptic region spanned by the data (see phase-space of $\varepsilon$-$\varepsilon^v$-$\dot{\varepsilon}^v$ in Fig.~\ref{Fig:viscoelastic_data}(d)). Consequently, the corners of this squared domain reflect the extrapolation of $f(\varepsilon, \varepsilon^v)$ and its derivative beyond the training dataset. Even when counting for these extrapolations, all the predictions have a very good agreement with the analytic values. The relative $L^2$ errors for $f$, $\sigma$, $\sigma^v$, $\psi$ and $f_v$ within the plotted range are $0.59\%$, $0.29\%$, $7.86\%$, $2.80\%$ and $2.40\%$, respectively.

\section{Example 3. Linear and nonlinear diffusion processes}
\label{Sec:Diffusion}

As a last set of examples, we apply the proposed material characterization strategy to two one-dimensional diffusion problems, namely, a linear and a nonlinear diffusion model. Following Sec.~\ref{Sec:Onsager_Diffusion}, we use the concentration $c$ and the flux $j$ as the state and process variables, respectively, and hence write the free energy density and dissipation potential densities as $f(c)$ and $\psi(c, j)$. In contrast to the two previous examples, $\psi$ here depends on both the state and the  process variable, and thus a Partially Input Convex Integrable Neural Network (PICINN) shall be used to discover it from data. Furthermore, these examples are characterized by a lack of uniqueness in $f$ and $\psi$, as discussed in Sec.~\ref{Sec:Onsager_Diffusion}, making them particularly interesting to analyze.


\subsection{Model description}
\label{Sec:Diffusion_Model}

The linear and nonlinear diffusion models considered are
\begin{align}
    \label{Eq:ZRP1_PDE}
    &\dot{c} = \frac{\partial^2 c}{\partial X^2} \quad \text{and} \\ \label{Eq:ZRP2_PDE}
    &\dot{c} = \frac{\partial }{\partial X}
    \left[ m(c) \frac{\partial \log (2m(c))}{\partial X} \right]
    \quad \text{with} \quad
    c(m) = \sqrt{2m} \frac{I_1\left(2\sqrt{2m}\right)}{I_0\left(2\sqrt{2m}\right)}, 
\end{align}
respectively, where $I_i$ are the modified Bessel functions of the first kind.

These two models arise, for instance, as the hydrodynamic limit of two symmetric zero range processes (ZRPs), that is, of  two particle jump processes on a lattice with different jump rate functions; for details see \citep{embacher2018computing}. For these specific processes, the free energy and dissipation potential densities can be analytically computed and read as
\begin{equation}
\label{Eq:f_psi_psitilde_ZRP1}
\begin{gathered}
    f(c) = \beta^{-1} \left( c \log c - c \right),
    \quad \text{and} \quad 
    \psi(c, j) = \frac{j^2}{2\beta c},
\end{gathered}
\end{equation}
for the linear model, and 
\begin{equation}
\label{Eq:f_psi_psitilde_ZRP2}
\begin{gathered}
    f(c) = \beta^{-1} \left[ c \log (2m(c)) 
    - \log I_0 \left(2 \sqrt{2m(c)} \right) \right],
    \quad \text{and} \quad 
    \psi(c, j) = \frac{j^2}{2\beta m(c)},
\end{gathered}
\end{equation}
for the nonlinear one. However, we recall, as discussed in Sec.~\ref{Sec:Onsager_Diffusion}, that there may be different physical processes, characterized by distinct free energy densities $f(c)$ and dissipation potential densities $\psi(c, j)$ that give rise to the same evolution equation. The only requirement is that the auxiliary function $\hat{\psi}(c,j) = \psi(c, j) / f''(c)$ is unique. Hence, only the auxiliary functions $\hat{\psi}(c, j) = j^2/2$ and $\hat{\psi}(c, j) = j^2/ \left[ 2 m'(c)\right]$ may be uniquely determined from data for these linear and nonlinear models.

\subsection{Data generation}

The diffusion models considered are simulated on the one-dimensional domain $[0,1]$ with periodic boundary conditions starting with an initial concentration profile $c(X,0) = 0.5 + 0.49 \sin \left( 4 \pi X \right)$. The domain is uniformly discretized into $N_X=99$ elements of length $\Delta X = 1/N_X$, and a constant time step $\Delta t$ is used for the temporal evolution. Then, denoting $c_i^n \coloneqq c(X_i, t^n)$ and $j_{i+\frac{1}{2}}^n \coloneqq j\left(\frac{X_i+X_{i+1}}{2}, t^n\right)$,
the linear diffusion model given by Eq.~\eqref{Eq:ZRP1_PDE} is solved by the forward time central space (FTCS) scheme \citep{lascaux1976lecture}
\begin{equation}
\label{Eq:ZRP1_PDE_discr}
    \frac{c_i^{n+1}-c_i^n}{\Delta t} 
    = \frac{c_{i+1}^n - 2c_i^n + c_{i-1}^n}{\Delta X^2}
    \quad \quad \text{and} \quad \quad 
    j_{i+\frac{1}{2}}^n 
    = \frac{c_{i+1}^n - c_i^n }{\Delta X},
\end{equation}
and the nonlinear model is evolved according to a conservative FTCS scheme
\begin{equation}
\label{Eq:ZRP2_PDE_discr}
\begin{gathered}
    \frac{c_i^{n+1}-c_i^n}{\Delta t} = \frac{ m_{i+\frac{1}{2}}^n 
    \left( \log 2m_{i+1}^n - \log 2m_{i}^n \right) - m_{i-\frac{1}{2}}^n     \left( \log 2m_{i}^n - \log 2m_{i-1}^n \right)}{\Delta X^2},\\
    \text{and} \quad \quad 
    j_{i+\frac{1}{2}}^n 
    = \frac{m_{i+\frac{1}{2}}^n 
    \left( \log 2m_{i+1}^n - \log 2m_{i}^n \right)}{\Delta X}.
\end{gathered}
\end{equation}
The total simulation time is $T = 0.025$, and the time steps $\Delta t = 2.55 \times 10^{-5}$ and $\Delta t =1.01 \times 10^{-5}$ are used to simulate the linear and nonlinear models, respectively.

\begin{figure}[h]
    \centering
    \includegraphics[width=\textwidth]{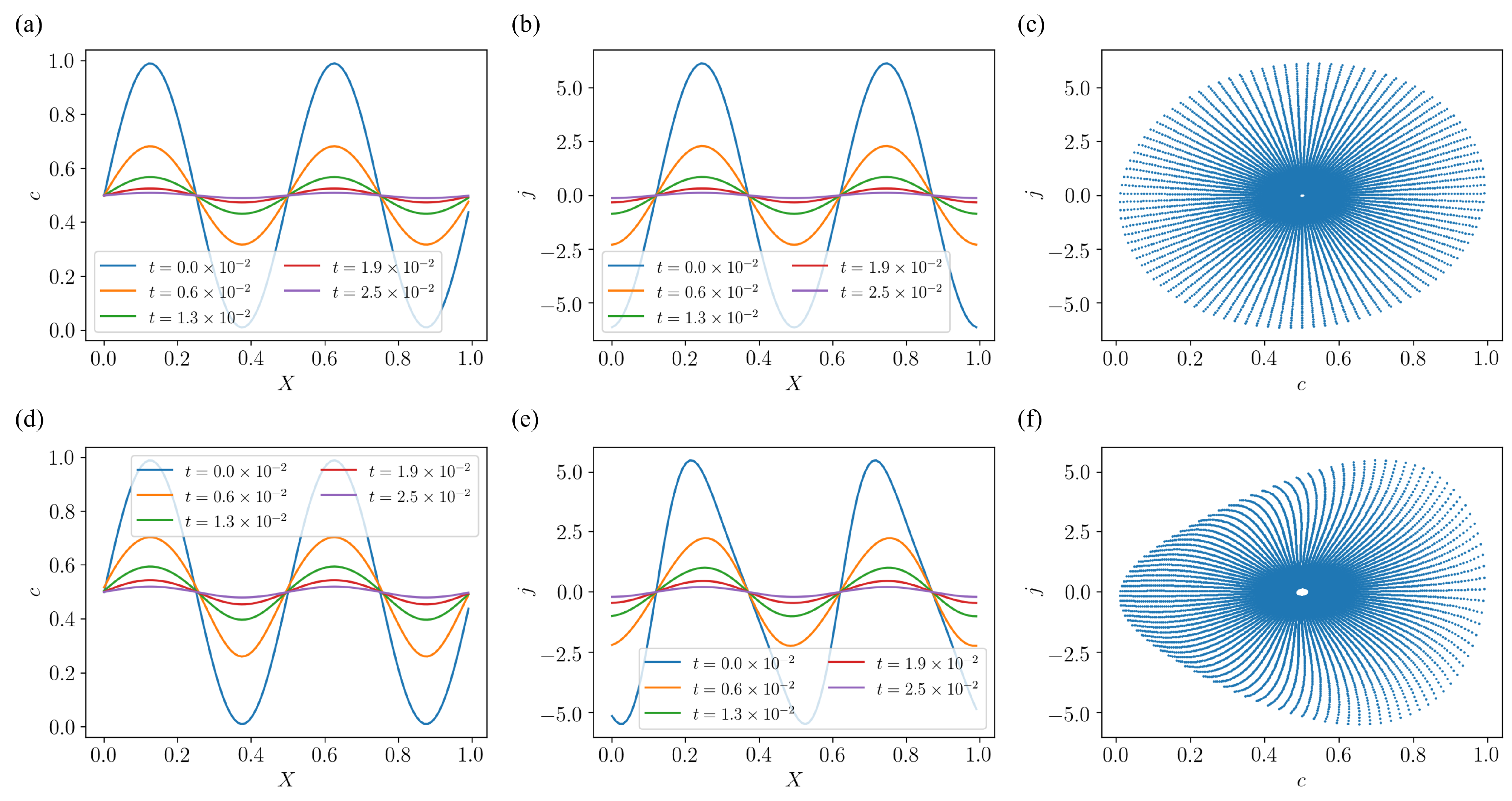}
    \caption{Dataset for (a-c) linear and (d-f) nonlinear diffusion models. Snapshots of (a,d) concentration fields and (b,e) flux fields. (c,f) Trajectories for all spacial points in the phase-space of $c$-$j$.}
    \label{Fig:diffusion_c_j_data}
\end{figure}

In order to avoid the dataset to be too large, only 201 snapshots (including the initial time) of $c$ and $j$ are outputted at a coarser time step for both models. Figures~\ref{Fig:diffusion_c_j_data}(a, b, d, e) show the evolution of the concentration fields $c(X)$ and flux $j(X)$ at various instances of time, and Figure~\ref{Fig:diffusion_c_j_data}(c, f) shows the distribution of this reduced dataset in the phase space of $c$-$j$. Here, as well, we randomly separate this data, so that $80\%$ is used for training and the remainder $20\%$ is used for testing.

\subsection{Networks architecture and loss function}
\label{Sec:Architecture_diffusion}

Although the free energy and dissipation potential densities cannot be uniquely determined from the data, we still consider independent NNs to approximate each of these potentials, as opposed to a single one for the auxiliary function $\hat{\psi}(c,j)$. This is because the convexity of $\hat{\psi}$ with respect to $j$ is not guaranteed as long as the convexity of $f(c)$ is unknown. Hence, we use an INN to represent the non-dimensional free energy density $\tilde{f}(\tilde{c})$, and a PICINN to represent the non-dimensional dissipation potential density $\tilde{\psi}(\tilde{c}, \tilde{j})$. The INN used has 2 hidden layers with 10 neurons for each layer, and the PICINN has 2 hidden layers with 10 neurons for each layer of the classical portion (shown as $\boldsymbol x_i$ in Fig.~\ref{Fig:NN}(c)) and 10 neurons for each layer of the convex portion (shown as $\boldsymbol y_i$ in Fig.~\ref{Fig:NN}(c)), i.e., 20 neurons in total per layer.

The dimensional potentials $f$ and $\psi$ are then computed by means of 
Eqs.~\eqref{Eq:rescale_fe} and \eqref{Eq:rescale_psi} from the output of the two networks. Since these potentials are not unique, the characteristic scales are simply set as
\begin{equation}
    f^* = 1,
\end{equation}
\begin{equation}
    \psi^* = \max \left( \left|j\right| \right),
\end{equation}
where the maximum is computed over the whole range of the input data.

The loss function $\mathcal{L}$ is  defined based on the discretized version of Eq.~\eqref{Eq:DiffKinetic} as,
\begin{equation}
\label{Eq:Loss_diffusion}
    \mathcal{L} 
    = \frac{1}{N_{PDE}} 
    \sum_{j=1}^{N_{PDE}} 
    \left| \frac{f'(c_{i+1}^n; \boldsymbol \theta_f)
    - f'(c_{i}^n; \boldsymbol \theta_f)}{\Delta X}
    + \psi_{,j} \left( c_i^n, j_{i+\frac{1}{2}}^n ; \boldsymbol \theta_{\psi} \right)  \right|^2.
\end{equation}
Since there is only one term in the loss function, no loss weight is required during the training process. Using the notation introduced in Sec.~\ref{Sec:Loss}, the input dataset $\{\boldsymbol{\xi}_{PDE}^j\}_{j=1}^{N_{PDE}}$ is defined as $\boldsymbol{\xi}_{PDE}^j = \left( c_i^n, c_{i+1}^n, j_{i+\frac{1}{2}}^n \right)$ with $N_{PDE} \leq N_X(N_T+1)$ and $j$ as a 1D index that denotes the flattened order of a subset of the 2D index $(i,n)$ with $i=0, \dots, N_X-1$ and $n = 0, \dots, n_T$.

\subsection{Training and results}
\label{Sec:Result_Diffusion}

\begin{figure}[h]
    \centering
    \includegraphics[width=0.8\textwidth]{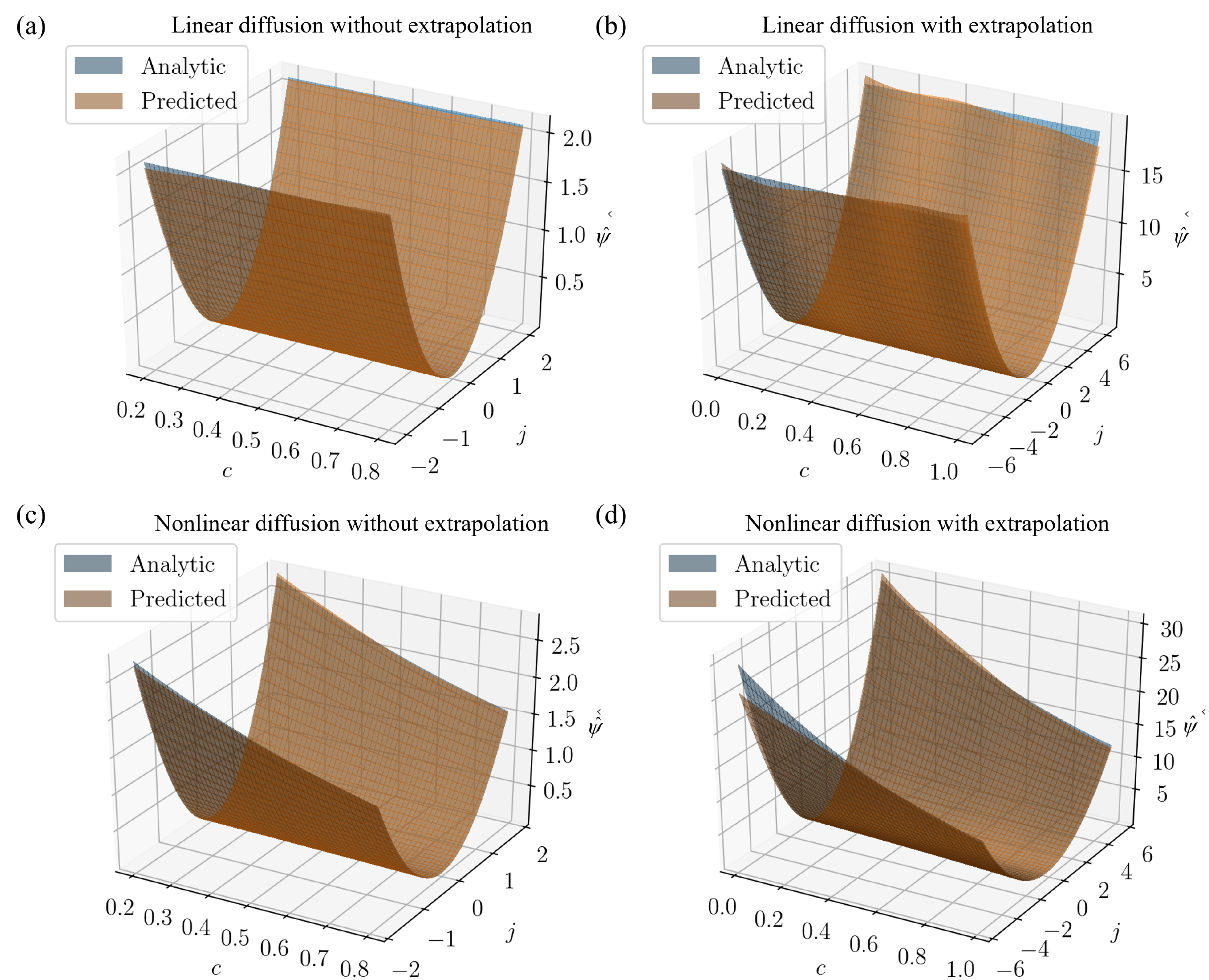}
    \caption{Training results. (a) and (b) are the predicted results of the linear diffusion model. (c) and (d) are the predicted results of the nonlinear diffusion model.}
    \label{Fig:diffusion_results}
\end{figure}

For both the linear and nonlinear diffusion examples, the models are trained for 12,000 epochs using an Adam optimizer with a learning rate of $8 \times 10^{-4}$. The training results are shown in Fig.~\ref{Fig:diffusion_results}. Here, subfigures (a) and (b) depict the auxiliary function $\hat{\psi}(c, j)$ for the linear diffusion model without and with extrapolation, and subfigures (c) and (d) show the corresponding results for the nonlinear example. Overall, the predictions exhibit a very good agreement with the analytical results in the region of available data, and only slightly larger differences are observed upon extrapolation. The relative $L^2$ errors of $\hat{\psi}$ for the results shown in Figs.~\ref{Fig:diffusion_results}(a)-(d) are $1.02\%$, $2.13\%$, $0.90\%$ and $4.75\%$ respectively.

\section{Conclusions}



%


In this paper, we designed neural networks, called Variational Onsager Neural Networks (VONNs), to discover non-equilibrium PDEs by learning the action density of the associated variational principle, namely, the free energy and the dissipation potential densities. By leveraging the variational structure, the proposed learning strategy only requires as input the state and process variables that describe the system and its evolution, while the operators of the ensuing PDEs and associated BCs will automatically follow from such choice and Onsager's variational principle. Furthermore, the proposed neural networks architecture to learn the potentials strongly enforces the second law of thermodynamics, hence guaranteeing the thermodynamic consistency of the learned evolution equations. For this, we combine the essence of Physics-Informed Neural Networks (PINNs), Integrable Deep Neural Networks (IDNNs) and Fully/Partial Input Convex Neural Networks (FICNNs/PICNNs), and further make use of the recently developed adaptative loss weight strategy to obtain a robust learning strategy with minimal user intervention.

We demonstrate this approach on a wide range of physical processes, including the phase transformation of a coiled-coil protein (characterized a non-convex free energy density), the one-dimensional approximation of the viscoelastic response of a three-dimensional model, and linear and nonlinear diffusion phenomena. In all cases, the free energy and dissipation potential densities are learned with great accuracy from spatio-temporal measurements of macroscopic observables, obtained from specific non-equilibrium processes. These results indicate that this new paradigm represents a promising and versatile avenue for learning the PDEs governing general non-equilibrium phenomena, as well as to construct physically based reduced-order models with guaranteed thermodynamic consistency.

Natural extensions of the current work include its integration with the VPINNs strategy to increase the computational efficiency, its application to higher-dimensional problems where the model is potentially unknown, and the addition of surface effects, of importance to sharp interface models. These will be the focus of future investigations.

\section{Code availability}
The codes used to generate the results here presented will be made available upon request to the corresponding author.

\section{Acknowledgement}

The authors gratefully acknowledge support from NSF CAREER Award, CMMI-2047506 (S.~H.~and C.~R.) and the U.S. Department of Education GAANN fellowship, P200A160282 (B.~C.). The authors further thank Prof.~Paris Perdikaris and Mr.~Yibo Yang for valuable discussions, particularly regarding the adaptive loss weights method here used to train the neural networks. 

\appendix

\section{Convexity for the PICINNs}
\label{App:Convex_FICINN}

In this section, we prove the conditions on the activation functions and weights that ensure the convexity of the PICINNs with respect to $\tilde{\mathbf{w}}$. Toward this goal, we denote the dependence of $\boldsymbol{y}_{i+1}$ on the input variables $\tilde{\mathbf{z}}$ and $\tilde{\mathbf{w}}$ with the function $\boldsymbol{y}_{i+1}(\tilde{\mathbf{z}}, \tilde{\mathbf{w}})$ while we denote the dependence of $\boldsymbol{y}_{i+1}$ on the previous layer $\boldsymbol{x}_i$ and $\boldsymbol{y}_i$ and the input variable $\tilde{\mathbf{w}}$ through the passthrough as $\hat{\boldsymbol{y}}_{i+1}(\boldsymbol{x}_i, \boldsymbol{y}_i, \tilde{\mathbf{w}})$, i.e.,
\begin{equation}
\label{Eq:PICINN_rewrite}
\begin{split}
    \boldsymbol{y}_{i+1} & = \boldsymbol{y}_{i+1}(\tilde{\mathbf{z}},\tilde{\mathbf{w}}) \\
    & = \hat{\boldsymbol{y}}_{i+1}(\boldsymbol{x}_i, \boldsymbol{y}_i, \tilde{\mathbf{w}}) 
    = g_i \left( \boldsymbol{Y}_i(\boldsymbol{x}_i, \boldsymbol{y}_i, \tilde{\mathbf{w}}) \right), \quad i = 0, 1, \dots, k,
\end{split}
\end{equation}
where $\boldsymbol{Y}_i(\boldsymbol{x}_i, \boldsymbol{y}_i, \tilde{\mathbf{w}})$ is given by (see Eq.~\eqref{Eq:PICINN})
\begin{equation}
    \boldsymbol{Y}_i(\boldsymbol{x}_i, \boldsymbol{y}_i, \tilde{\mathbf{w}})
    = \boldsymbol{W}_i^y \left[ \boldsymbol{y}_i \circ 
    g_i^{yx} \left( \boldsymbol{W}_i^{yx} \boldsymbol{x}_i + \boldsymbol{b}_i^{yx} \right) \right]
    + \boldsymbol{W}_i^w \left[ \tilde{\mathbf{w}} \circ 
    \left( \boldsymbol{W}_i^{wx} \boldsymbol{x}_i + \boldsymbol{b}_i^{wx} \right) \right]
    + \boldsymbol{W}_i^x \boldsymbol{x}_i + \boldsymbol{b}_i.
\end{equation}

The first order derivative of $\boldsymbol{y}_{i+1}$ with respect to $\tilde{\mathbf{w}}$ can be readily computed as,
\begin{equation}
\label{Eq:PICINN_dw}
\begin{split}
    \left( \frac{\partial \boldsymbol{y}_{i+1}}{\partial \tilde{\mathbf{w}}} \right)_{\tilde{\mathbf{z}}} 
    & = \left( \frac{\partial \hat{\boldsymbol{y}}_{i+1}}{\partial \boldsymbol{y}_i} \right)_{\boldsymbol{x}_i, \tilde{\mathbf{w}}}
    \left( \frac{\partial \boldsymbol{y}_i}{\partial \tilde{\mathbf{w}}} \right)_{\tilde{\mathbf{z}}}
    + \left( \frac{\partial \hat{\boldsymbol{y}}_{i+1}}{\partial \tilde{\mathbf{w}}} \right)_{\boldsymbol{x}_i,\boldsymbol{y}_i} \\
    & = g_i' \left( \boldsymbol{Y}_i(\boldsymbol{x}_i, \boldsymbol{y}_i, \tilde{\mathbf{w}}) \right)
    \left[ \boldsymbol{W}_i^y g_i^{yx} \left( \boldsymbol{W}_i^{yx} \boldsymbol{x}_i + \boldsymbol{b}_i^{yx} \right) 
    \left( \frac{\partial \boldsymbol{y}_i}{\partial \tilde{\mathbf{w}}} \right)_{\tilde{\mathbf{z}}} + \boldsymbol{W}_i^w \left( \boldsymbol{W}_i^{wx} \boldsymbol{x}_i + \boldsymbol{b}_i^{wx} \right) \right],
\end{split}
\end{equation}
where we have used the fact that $\boldsymbol{x}_i$ does not depend on $\tilde{\mathbf{w}}$. 

Similarly, the second order derivative of $\boldsymbol{y}_{i+1}$ with respect to $\tilde{\mathbf{w}}$ can be written as,
\begin{equation}
\label{Eq:PICINN_dwdw}
\begin{split}
    \left( \frac{\partial^2 \boldsymbol{y}_{i+1}}{\partial \tilde{\mathbf{w}}^2} \right)_{\tilde{\mathbf{z}}} 
    & = g_i'' \left( \boldsymbol{Y}_i(\boldsymbol{x}_i, \boldsymbol{y}_i, \tilde{\mathbf{w}}) \right)
    \left[ \boldsymbol{W}_i^y g_i^{yx} \left( \boldsymbol{W}_i^{yx} \boldsymbol{x}_i + \boldsymbol{b}_i^{yx} \right) 
    \left( \frac{\partial \boldsymbol{y}_i}{\partial \tilde{\mathbf{w}}} \right)_{\tilde{\mathbf{z}}} + \boldsymbol{W}_i^w \left( \boldsymbol{W}_i^{wx} \boldsymbol{x}_i + \boldsymbol{b}_i^{wx} \right) \right]^2 \\
    & \quad 
    +  g_i' \left( \boldsymbol{Y}_i(\boldsymbol{x}_i, \boldsymbol{y}_i, \tilde{\mathbf{z}}) \right)
    \left[ \boldsymbol{W}_i^y g_i^{yx} \left( \boldsymbol{W}_i^{yx} \boldsymbol{x}_i + \boldsymbol{b}_i^{yx} \right) \right]
    \left( \frac{\partial^2 \boldsymbol{y}_i}{\partial \tilde{\mathbf{w}}^2} \right)_{\tilde{\mathbf{z}}}.
\end{split}
\end{equation}

In view of this recurrence relation, partial convexity of the PICINN with respect to $\tilde{\mathbf{w}}$, i.e., $\left( \frac{\partial^2 \boldsymbol{y}_{k+1}}{\partial \tilde{\mathbf{w}}^2} \right)_{\tilde{\mathbf{z}}} \geq 0$, is ensured if $g_i''(\cdot)\geq 0$ and $g_i'(\cdot)\geq 0$ (i.e., $g_i(\cdot)$ is convex and non-decreasing), $\boldsymbol{W}_i^y \geq 0$ and $g_i^{yx}(\cdot) \geq 0$.


\section{On sampling bias and adaptative loss weights }
\label{App:PhaseTrans_Coarser_t_Constant_Weights}

This section provides two additional sets of training results for the first example on phase transformations, discussed in Sec.~\ref{Sec:Model_PhaseTrans}. These are aimed at highlighting the importance of proper sample selection and the adaptative loss weights on obtaining accurate results.

\begin{figure}[h]
    \centering
    \includegraphics[width=0.8\textwidth]{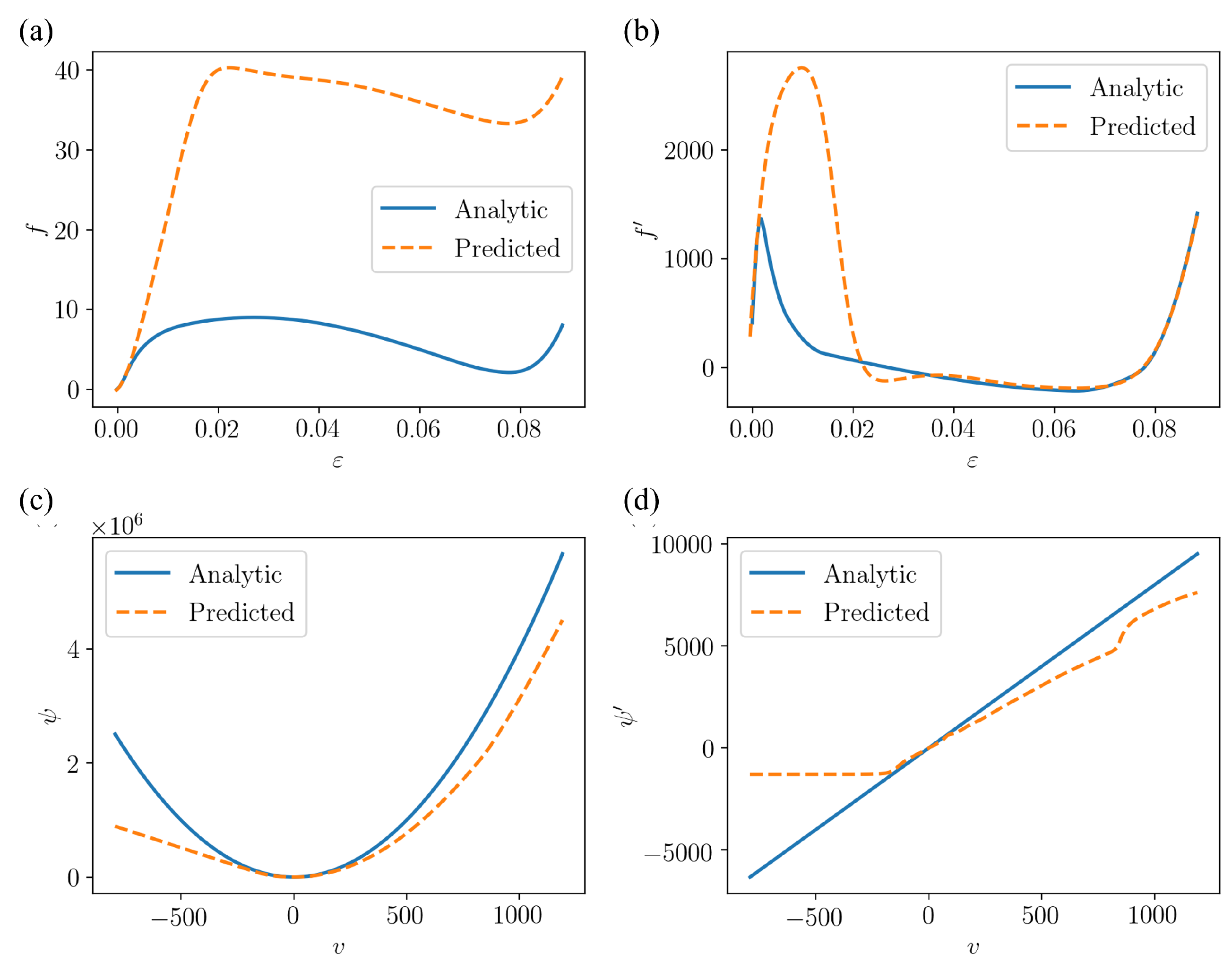}
    \caption{Training results using the data that was selected uniformly in time for (a) the free energy density $f(\varepsilon)$, (b) the stress $f'(\varepsilon)$, (c) the dissipation potential density $\psi(v)$ and (d) the viscous force $\psi'(v)$.}
    \label{Fig:equispaced_protein_prediction}
\end{figure}

As a first case, we consider uniform spatio-temporal data, with $N_X=150$ and $\Delta t_{coarse} = 3000\Delta t = 2.7 \times 10^{-5}$ ns, and use the adaptative loss weight strategy to calculate the loss function. Similarly to the results shown in the narrative, the model is trained for $30,000$ epochs using an Adam optimizer with a learning rate of $10^{-4}$. The training results for the free energy and dissipation potential densities and their derivatives are shown in Fig.~\ref{Fig:equispaced_protein_prediction}. Although much more data is here used in the training process compared to the results of Fig.~\ref{Fig:protein_prediction} (here, $N_{BC} = 1038 \times 80\% \simeq 830$ and $N_{PDE} = 1038 \times 150 \times 80\% = 124,560$), the predictions are markedly distinct from the analytical functions. This can be easily explained by means of Fig.~\ref{Fig:protein_data_compare}, where the boundary data (traction $\bar{t}$ versus boundary strain $\varepsilon_{BC}$) obtained from a uniform spatio-temporal sampling strategy is compared to the data selected on the basis of phase-space uniformity. The former data mainly concentrates at the two ends of the strain range, while the middle part, associated with the free energy barrier, is rarely captured. This is consistent with the results for $f'(\varepsilon)$ shown in Fig.~\ref{Fig:equispaced_protein_prediction}(b), where the predictions only agree with the analytic values where the data is available.

\begin{figure}[h]
    \centering
    \includegraphics[width=0.8\textwidth]{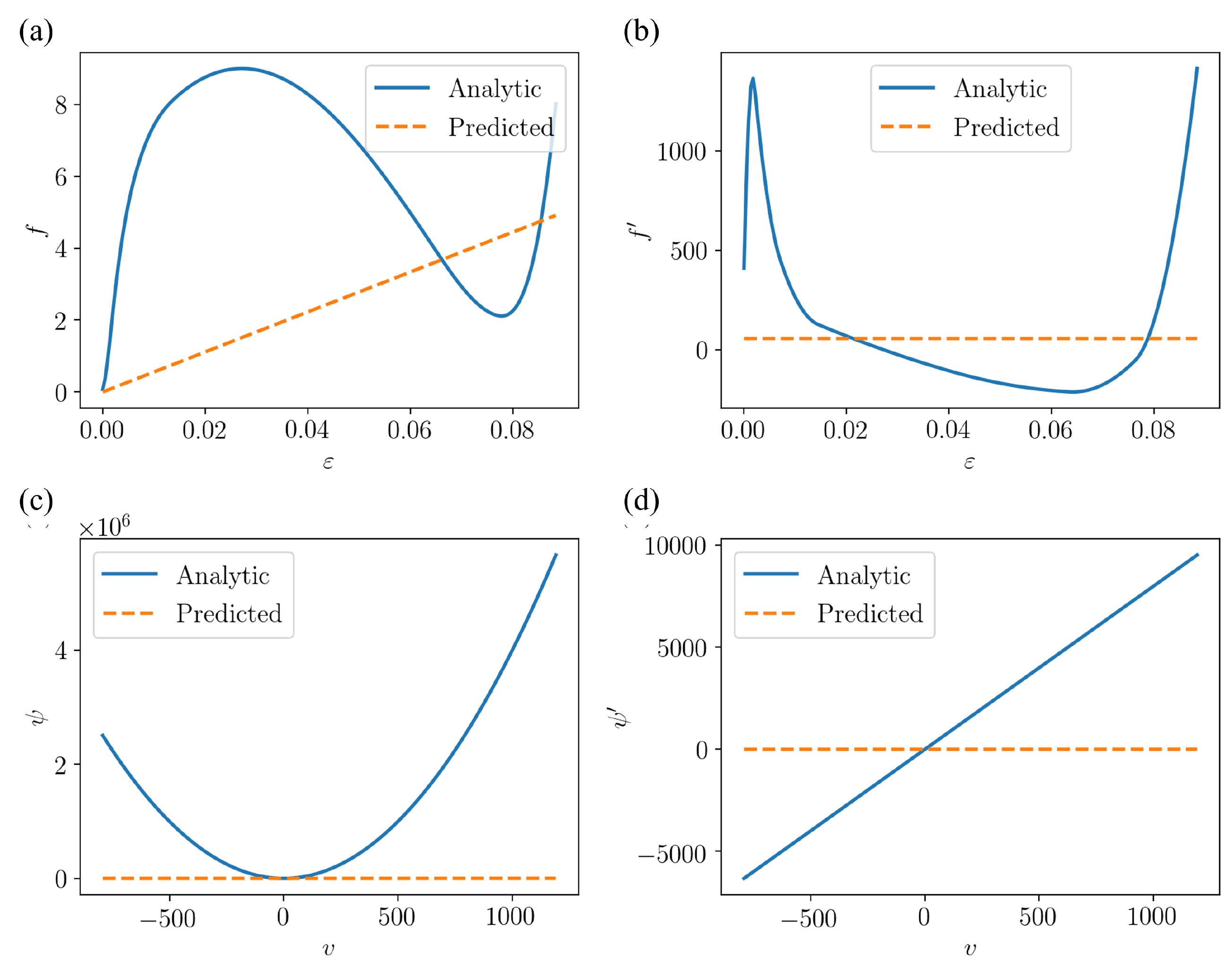}
    \caption{Training results using constant loss weights estimated by dimensional analysis for (a) the free energy density $f(\varepsilon)$, (b) the stress $f'(\varepsilon)$, (c) the dissipation potential density $\psi(v)$ and (d) the viscous force $\psi'(v)$.}
    \label{Fig:static_protein_prediction}
\end{figure}

Next, we demonstrate the importance of using the adaptive loss weights method described in Sec.~\ref{Sec:Loss} to obtain accurate results. For this, we use the same training dataset as that used in Sec.~\ref{Sec:Result_PhaseTrans}, but estimate instead the loss weights to be used in the loss function,  Eq.~\eqref{Eq:Loss_PhaseTrans}, based on dimensional analysis as
\begin{equation}
    \alpha_{PDE} = \left(\frac{L}{\sigma_{\bar{t}}}\right)^2,
\end{equation}
\begin{equation}
    \alpha_{BC} =\frac{1}{\sigma_{\bar{t}}^2}.
\end{equation}
Here, $\sigma_{\bar{t}}$ is the standard deviation of the data for the applied traction $\bar{t}$. The model is then trained for $30,000$ epochs using an Adam optimizer and a learning rate of $10^{-4}$, similarly to the results of the main text in the narrative. The results are shown in Fig.~\ref{Fig:static_protein_prediction}, where there is not even a qualitative agreement with the analytical results for any range of the data. Actually, the trained model appears to be trapped in a local minima, where $f'(\varepsilon)$ and $\psi'(v)$ are approximately zero, and hence $\mathcal{L}_{PDE} \approx 0$. Although this problem may be resolved by fine-tuning the loss weights manually, together with the learning rate and training epochs, this process can be time consuming and hard to realize when there are many loss weights and the analytical results are unknown.


\section{From 3D to 1D compressible viscoelasticity}
\label{App:3D_to_1D}

We here derive the 1D viscoelastic evolution equations form the 3D model, given by Eqs.~\eqref{Eq:StressStrain} and \eqref{Eq:ViscousStrainRelaxation}, under one-dimensional loading conditions
\begin{equation}
    \boldsymbol{\sigma} = 
    \begin{pmatrix}
    \sigma & 0 & 0 \\
    0 & 0 & 0 \\
    0 & 0 & 0 \\
    \end{pmatrix}
\end{equation}
and the assumptions that $\text{tr}\left(\pmb{\varepsilon}^{v \alpha}(\mathbf{X},0)\right)=0$ and $\varepsilon_{ij}^{v \alpha}(\mathbf{X},0)=0$, for $i\neq j$.

First, we note by taking trace on both sides of Eq.~\eqref{Eq:ViscousStrainRelaxation} that 
\begin{equation}
    \tau_{\alpha} \frac{d}{dt} \text{tr} \left(\pmb{\varepsilon}^{v \alpha} \right)
    = - \text{tr} \left(\pmb{\varepsilon}^{v \alpha} \right),
\end{equation}
and, hence, $\pmb{\varepsilon}^{v \alpha}$ is traceless at all times, since it is assumed to be traceless at $t=0$.


Next, taking trace on both sides of Eq.~\eqref{Eq:StressStrain}, and noting the one-dimensional loading assumption,
\begin{equation}
    \text{tr}\left( \boldsymbol{\sigma} \right) =\sigma 
    = 3 K \text{tr}\left(\pmb{\varepsilon} \right). 
\end{equation}
The 3D constitutive relation given by Eq.~\eqref{Eq:StressStrain} then reads as
\begin{equation}
    \left\{ \begin{aligned}
    & K \text{tr}(\pmb{\varepsilon}) + 
    2G \varepsilon_{11}^{dev} 
    + \sum_{\alpha=1}^n 2G_{v\alpha} \left( \varepsilon_{11}^{dev} - \varepsilon_{11}^{v \alpha} \right) 
    = \sigma \\
    & K \text{tr}(\pmb{\varepsilon}) + 
    2G \varepsilon_{22}^{dev} 
    + \sum_{\alpha=1}^n 2G_{v\alpha} \left( \varepsilon_{22}^{dev} - \varepsilon_{22}^{v \alpha} \right) 
    = 0 \\
    & K \text{tr}(\pmb{\varepsilon})
    + 2G \varepsilon_{33}^{dev} 
    + \sum_{\alpha=1}^n 2G_{v\alpha} \left( \varepsilon_{33}^{dev} - \varepsilon_{33}^{v \alpha} \right)
    = 0 \\
    & 2G \varepsilon_{ij}
    + \sum_{\alpha=1}^n 2G_{v\alpha} \left( \varepsilon_{ij} - \varepsilon_{ij}^{v \alpha} \right)
    = 0, 
    \quad \quad \text{for } i \neq j,
    \end{aligned}  \right.
\end{equation}
where we have used the fact that $\varepsilon_{ij}^{dev} = \varepsilon_{ij}$ for $i \neq j$. From the last equation, one can easily obtain the off-diagonal components of $\pmb{\varepsilon}$, i.e., 
\begin{equation}
    \varepsilon_{ij}
    = \frac{\sum_{\alpha=1}^n G_{v\alpha} \varepsilon_{ij}^{v \alpha}}{G + \sum_{\alpha=1}^n G_{v\alpha}}
    , \quad \quad \text{for } i \neq j.
\end{equation}
This equation, combined with Eq.~\eqref{Eq:ViscousStrainRelaxation}, indicates that $\varepsilon^{v \alpha}_{ij} (\mathbf{X}, t)\equiv 0$ for $i \neq j$ for all times as these are assumed zero at $t=0$.


Furthermore, by symmetry, it is immediate that $\varepsilon_{22} = \varepsilon_{33}$ and $\varepsilon_{22}^{v \alpha} = \varepsilon_{33}^{v \alpha}$, and since $\pmb{\varepsilon}^{dev}$ and $\pmb{\varepsilon}^{v \alpha}$ are traceless, then $\varepsilon_{22}^{dev} = \varepsilon_{33}^{dev} = - \varepsilon_{11}^{dev} / 2$ and $\varepsilon_{22}^{v \alpha} = \varepsilon_{33}^{v \alpha} = - \varepsilon_{11}^{v \alpha} / 2$. Denoting  for simplicity $\varepsilon = \varepsilon_{11}$ and  $\varepsilon^{v \alpha} = \varepsilon_{11}^{v \alpha}$,  the deviatoric part of strain can then be expressed as, 
\begin{equation}
    \varepsilon_{11}^{dev} 
    = \varepsilon 
    - \frac{1}{3} \text{tr} (\pmb{\varepsilon})
    = \varepsilon - \frac{\sigma}{9K}
\end{equation}
The constitutive equations for viscoelastic materials with 1D loading can thus be written as,
\begin{equation}
\begin{split}
    \sigma 
    & = K \text{tr}(\pmb{\varepsilon}) 
    + 2G \varepsilon_{11}^{dev} 
    + \sum_{\alpha=1}^n 2G_{v\alpha} \left( \varepsilon_{11}^{dev} - \varepsilon^{v \alpha} \right) \\
    & = \frac{\sigma}{3}
    + 2G \varepsilon 
    + \sum_{\alpha=1}^n 2G_{v\alpha} \left( \varepsilon - \varepsilon^{v \alpha} \right)
    - \left( 2G + \sum_{\alpha=1}^n 2G_{v\alpha} \right) \frac{\sigma}{9K}
\end{split}
\end{equation}
\begin{equation}
    \tau_{\alpha} \dot{\varepsilon}^{v \alpha} 
    = \varepsilon_{11}^{dev} - \varepsilon^{v \alpha}
    = \varepsilon - \frac{\sigma}{9K}
    - \varepsilon^{v \alpha}.
\end{equation}
After further simplification, these equations can be expressed as
\begin{equation} \label{Eq:ConstRelation1D}
    \sigma = E_{1D} \varepsilon 
    + \sum_{\alpha=1}^n E_{1D\alpha} \left( \varepsilon - \varepsilon^{v \alpha} \right),
\end{equation}
with
\begin{equation}
    E_{1D} = \frac{3G}{\theta},
    \quad \quad 
    E_{1D\alpha} = \frac{3G_{v\alpha}}{\theta}
    \quad \quad \text{and} \quad \quad
    \theta = 1 + 
    \frac{G + \sum_{\alpha=1}^n G_{v\alpha}}{3K},
\end{equation}
and 
\begin{equation} \label{Eq:EvolEq1D}
\begin{split}
    \tau_{\alpha} \dot{\varepsilon}^{v \alpha} 
    & =  \varepsilon - \varepsilon^{v \alpha}
    - \frac{1}{9K} \left[  E_{1D} \varepsilon 
    + \sum_{\gamma=1}^n E_{1D\gamma} \left( \varepsilon - \varepsilon^{v \gamma} \right) \right]. 
\end{split}
\end{equation}

Now, we look at the variational characterization of these equations and show how these result from Onsager's variational principle under the one-dimensional loading assumption. First, we recall the 3D version of the free energy and dissipation potential density, i.e.,
\begin{equation}
    f_{3D} \left( \pmb{\varepsilon}, \{\pmb{\varepsilon}^{v\alpha}\}_{\alpha=1}^n \right)
    = \frac{1}{2} K \left( \text{tr} \pmb{\varepsilon} \right)^2
    + G \pmb{\varepsilon}^{dev} : \pmb{\varepsilon}^{dev}
    + \sum_{\alpha=1}^n G_{v\alpha}
    \left( \pmb{\varepsilon}^{dev} - \pmb{\varepsilon}^{v \alpha} \right) 
    : \left( \pmb{\varepsilon}^{dev} - \pmb{\varepsilon}^{v \alpha} \right) 
\end{equation}
\begin{equation}
    \psi_{3D} \left( \{\dot{\pmb{\varepsilon}}^{v\alpha}\}_{\alpha=1}^n \right)
    = \sum_{\alpha=1}^n \frac{1}{2} \eta_{\alpha}
    \dot{\pmb{\varepsilon}}^{v\alpha}
    : \dot{\pmb{\varepsilon}}^{v\alpha},
\end{equation}
where $\eta_\alpha = 2G_{v\alpha} \tau_\alpha$. 
The 3D equilibrium equations and evolution equation for the internal variables can be written by means of these potentials as
\begin{equation}
    \rho \textbf{a} 
    = \nabla \cdot \pmb \sigma 
    = \nabla \cdot 
    \left(  \frac{\partial f_{3D}}{\partial \pmb{\varepsilon}} \right)
\end{equation}
\begin{equation}
    \frac{\partial \psi_{3D}}{\partial \dot{\pmb{\varepsilon}}^{v\alpha}}
    + \frac{\partial f_{3D}}{\partial \pmb{\varepsilon}^{v\alpha}}  = 0
\end{equation}

The one-dimensional free energy density may then be obtained from the 3D one by inserting the  one-dimensional loading ansatz, i.e.
\begin{equation}
\begin{split}
    & f_{1D} \left( \varepsilon, \{\varepsilon^{v\alpha}\}_{\alpha=1}^n \right) 
    \coloneqq  f_{3D} \left( \pmb{\varepsilon}\left(\varepsilon, 
    \left\{ \varepsilon^{v\alpha} \right\}_{\alpha=1}^n \right), \left\{\pmb{\varepsilon}^{v\alpha} \left( \varepsilon^{v\alpha} \right) \right\}_{\alpha=1}^n \right) \\
    & = \frac{1}{2} K \left( \text{tr} \pmb{\varepsilon} \right)^2
    + G \pmb{\varepsilon}^{dev} : \pmb{\varepsilon}^{dev}
    + \sum_{\alpha=1}^n G_{v\alpha}
    \left( \pmb{\varepsilon}^{dev} - \pmb{\varepsilon}^{v \alpha} \right) 
    : \left( \pmb{\varepsilon}^{dev} - \pmb{\varepsilon}^{v \alpha} \right) \\
    & = \frac{1}{2} K \left(  \text{tr} \pmb{\varepsilon}  \right)^2
    + \frac{3}{2} G \left( \varepsilon_{11}^{dev} \right)^2
    + \sum_{\alpha=1}^n \frac{3}{2} G_{v\alpha}
    \left( \varepsilon_{11}^{dev} - \varepsilon^{v \alpha} \right)^2 \\
    & = \frac{1}{2} K \left( \frac{\sigma}{3K} \right)^2
    + \frac{3}{2} G \left( \varepsilon - \frac{\sigma}{9K} \right)^2
    + \sum_{\alpha=1}^n \frac{3}{2} G_{v\alpha}
    \left( \varepsilon - \frac{\sigma}{9K} 
    - \varepsilon^{v \alpha} \right)^2 \\
    & = \frac{3}{2}G \varepsilon^2
    + \sum_{\alpha=1}^n \frac{3}{2} G_{v\alpha} \left( \varepsilon - \varepsilon^{v \alpha} \right)^2
    - \frac{\theta }{18 K} 
    \left[ E_{1D} \varepsilon + \sum_{\alpha=1}^n E_{1D\alpha} \left( \varepsilon - \varepsilon^{v \alpha} \right) \right]^2 \\
    & = \frac{1}{2} \theta E_{1D} \varepsilon^2
    + \sum_{\alpha=1}^n \frac{1}{2} \theta E_{1D\alpha} \left( \varepsilon - \varepsilon^{v \alpha} \right)^2
    -  \frac{\theta }{18 K} 
    \left[ E_{1D} \varepsilon + \sum_{\alpha=1}^n E_{1D\alpha} \left( \varepsilon - \varepsilon^{v \alpha} \right) \right]^2.
\end{split}
\end{equation}
The corresponding derivatives are given by,
\begin{equation}
\begin{split}
    \frac{\partial f_{1D}}{\partial \varepsilon}
    & = E_{1D} \varepsilon + \sum_{\alpha=1}^n E_{1D\alpha} \left( \varepsilon - \varepsilon^{v\alpha} \right)
    = \sigma,
\end{split}
\end{equation}
\begin{equation}
\begin{split}
    \frac{\partial f_{1D}}{\partial \varepsilon^{v\alpha}}
    & = - \theta E_{1D\alpha} \left( \varepsilon - \varepsilon^{v \alpha} \right)
    + \frac{\theta E_{1D\alpha}}{9K}
    \left[ E_{1D} \varepsilon + \sum_{\gamma=1}^n E_{1D\gamma} \left( \varepsilon - \varepsilon^{v \alpha} \right) \right] \\
    & = - 3 G_{v\alpha} \left( \varepsilon - \varepsilon^{v \alpha} \right)
    + \frac{G_{v\alpha}}{3K}
    \sigma.
\end{split}
\end{equation}
Similarly, the one-dimensional dissipation potential density can be obtained from the 3D density as
\begin{equation}
    \psi_{1D} \left( \{\dot{\varepsilon}^{v\alpha}\}_{\alpha=1}^n \right)\coloneqq\psi_{3D} \left( \{\dot{\pmb{\varepsilon}}^{v\alpha} (\dot{\varepsilon}^{v\alpha}) \}_{\alpha=1}^n \right)
    = \sum_{\alpha=1}^n \frac{1}{2} \eta_{\alpha} \dot{\pmb{\varepsilon}}^{v\alpha} : \dot{\pmb{\varepsilon}}^{v\alpha} 
    = \sum_{\alpha=1}^n \frac{3}{4} \eta_{\alpha} \left( \dot{\varepsilon}^{v\alpha} \right)^2
    = \sum_{\alpha=1}^n \frac{1}{2} \eta_{1D\alpha} \left( \dot{\varepsilon}^{v\alpha} \right)^2,
\end{equation}
where $\eta_{1D\alpha} = \frac{3}{2}\eta_{\alpha}$, and the corresponding derivative is given by,
\begin{equation}
    \frac{\partial \psi_{1D}}{\partial \dot{\varepsilon}^{v\alpha}}
    = \eta_{1D\alpha} \dot{\varepsilon}^{v\alpha}
    = 3 G_{v\alpha} \tau_{\alpha} \dot{\varepsilon}^{v\alpha}.
\end{equation}
It may then be readily observed that the 1D equilibrium equation and evolution equation for the internal variable, resulting from Eqs.~\eqref{Eq:ConstRelation1D} and \eqref{Eq:EvolEq1D}, can be equivalently written by means of the 1D free energy and dissipation potentials, following Onsager's variational principle, i.e.,
\begin{equation}
    \rho a 
    = \frac{\partial \sigma}{\partial X} 
    = \frac{\partial}{\partial X}
    \left(  \frac{\partial f_{1D}}{\partial \varepsilon} \right)
\end{equation}
\begin{equation}
    \frac{\partial \psi_{1D}}{\partial \dot{\varepsilon}^{v\alpha}}
    + \frac{\partial f_{1D}}{\partial \varepsilon^{v\alpha}} = 0.
\end{equation}
Similarly, the traction boundary conditions can be written as,
\begin{equation}
    \left. \sigma \right|_{boundary} 
    = \left. \frac{\partial f_{1D}}{\partial \varepsilon} \right|_{boundary} 
    = \bar{t}.
\end{equation}


\end{document}